\documentclass[12pt,preprint]{aastex}
\renewcommand{\cite}{\citealp}
\newcommand{\rrl}{{RR~Lyrae}}

\shorttitle{Variable Stars in Fornax~4}
\shortauthors{Greco et al.}

\begin{document}

\title{Variable Stars in the Fornax dSph Galaxy. I. The Globular Cluster Fornax~4\altaffilmark{1}}

\author{
Claudia Greco,\altaffilmark{2,3,4} 
Gisella Clementini,\altaffilmark{2}
M\'arcio Catelan,\altaffilmark{5}
Enrico V. Held,\altaffilmark{6}
Ennio Poretti,\altaffilmark{7}
Marco Gullieuszik,\altaffilmark{6,8} 
Marcella Maio,\altaffilmark{2}
%Luca Dell'Arciprete,\altaffilmark{8}
%Luca Rizzi,\altaffilmark{9}
Armin Rest,\altaffilmark{9}
Nathan De Lee,\altaffilmark{10}
Horace A. Smith,\altaffilmark{10}
Barton J. Pritzl\altaffilmark{11}
}

\altaffiltext{1}{This paper includes data gathered with the  
%Based on data collected at the 
6.5-m Magellan telescope located at 
Las Campanas Observatory, Chile, and with the 4-m Blanco telescope 
at the Cerro Tololo Inter-American Observatory, Chile.}

\altaffiltext{2}{INAF, Osservatorio Astronomico di
Bologna, via Ranzani 1, I-40127 Bologna, Italy;
(claudia.greco, gisella.clementini)@oabo.inaf.it}

\altaffiltext{3}{Dipartimento di Astronomia, Universit\`a di
Bologna, via Ranzani 1, I-40127 Bologna, Italy}

\altaffiltext{4}{Marco Polo Fellow at the Pontificia Universidad Cat\'olica
 de Chile, Departamento de Astronom\'{\i}a y Astrof\'{\i}sica}

\altaffiltext{5}{Departamento de Astronom\'{\i}a y Astrof\'{\i}sica, 
Pontificia Universidad Cat\'olica de Chile, Av. Vicu\~na Mackenna 4860,
782-0436 Macul, Santiago, Chile; mcatelan@astro.puc.cl}

\altaffiltext{6}{INAF, Osservatorio Astronomico di
Padova, vicolo dell'Osservatorio 5, I-35122 Padova, Italy;
(enrico.held, marco.gullieuszik)@oapd.inaf.it}

\altaffiltext{7}{INAF, Osservatorio Astronomico di
Brera, via E. Bianchi 46, 23807 Merate, Italy;
ennio.poretti@brera.inaf.it}

\altaffiltext{8}{Dipartimento di Astronomia, Universit\`a di
Padova, vicolo dell'Osservatorio 2, I-35122 Padova, Italy}

\altaffiltext{9}{Cerro Tololo Inter-American Observatory, Casilla 603, La Serena, Chile;
arest@ctio.noao.edu}

\altaffiltext{10}{Department of Physics and Astronomy, Michigan State University, East Lansing, MI 48824-2320, USA;
(smith,delee)@pa.msu.edu}

\altaffiltext{11}{Macalester College, 1600 Grand Avenue, Saint Paul, MN 55105, USA;
pritzl@macalester.edu}

%-----------------------------------------------------------------

\begin{abstract}

Variable stars have been identified for the first time in Fornax~4, 
the 
%youngest of the Fornax dwarf spheroidal galaxy (dSph) 
globular cluster located 
%in the central region of 
near the center of the Fornax dwarf spheroidal galaxy.
By applying  the  image subtraction technique to $B,V$ time series photometry
obtained with the MagIC camera of the 6.5-m Magellan/Clay telescope and with the wide field imager 
of the 4-m Blanco/CTIO telescope, we detected
27 RR Lyrae stars (22 fundamental mode, 3 first overtone, and 2 double-mode pulsators)
in a $2.4^{\prime}\times 2.4^{\prime}$ area centered on Fornax~4. 
The average and minimum periods of the {\it ab-}type RR Lyrae stars, 
$\langle P_{ab}\rangle = 0.594$~d and 
$P_{ab,min}=0.5191$~d, respectively, as well as the revised position
of the cluster in the horizontal branch type--metallicity plane, all
consistently point to an Oosterhoff-intermediate status for the cluster,
unlike what is seen for the vast majority of Galactic globular clusters,
but in agreement with previous indications for the other globular clusters 
in Fornax.

The average apparent magnitude of the RR Lyrae stars located within 
$30\arcsec$ from the cluster center is $\langle V(RR)\rangle=21.43 \pm 0.03$~mag
($\sigma=0.10$~mag, average on 12 stars), leading to a true distance modulus of
$\mu_0=20.64 \pm 0.09$~mag or $\mu_0=20.53 \pm 0.09$~mag, depending on whether 
a low (${\rm [Fe/H]}=-2.0$) or a moderately high (${\rm [Fe/H]}=-1.5$) metallicity  
is adopted. 

%if we adopt for the cluster the metal abundance of Buonanno et al. (1999, 
%${\rm [Fe/H]}=-2.0$), or 
% $\mu_0=20.53 \pm 0.09$~mag, if ${\rm [Fe/H]}=-1.5$, as derived by Strader et al. (2003),
% for an assumed reddening $E(\bv)=0.10$~mag.\\

\end{abstract}


%-----------------------------------------------------------------
\section{Introduction}
This is the first in a series of papers devoted to the detailed and 
comprehensive study
of the variable star population in the field and globular clusters of
the Fornax dwarf spheroidal galaxy (dSph) (The Fornax Project, Clementini et al. 2006), encompassing the galaxy's classical instability
strip from the Anomalous Cepheid ($V \sim 19$~mag) to the Dwarf Cepheid
($V \sim 25$~mag) region (Poretti et al. 2006, 2007). Here, we report on results from the first 
variability study in Fornax~4 (For~4), the most compact and centrally located of the Fornax globular clusters
(GC's), 
for which it has been suggested that it might actually be the nucleus of the Fornax dSph galaxy
(Hardy 2002; Strader et al. 2003). In the following papers of
this series, we will report on our study of the variable stars identified in the  
Fornax clusters \#2, 3 and 5, as well as 
in the survey of about a 1~deg$^2$ of the galaxy field.

Fornax and Sagittarius are the only dSph's known to host globular clusters (GC's). The 
Fornax dSph GC system contains 5 GC's  (Hodge 1961, 1965, 1969).
Buonanno et al. (1998, 1999) published color-magnitude diagrams (CMDs) of the 
Fornax GC's based on {\it Hubble Space Telescope} (HST)
 observations from which they estimated the clusters' relative ages and metallicities.
They concluded that 
For~4\  differs both in horizontal branch (HB) morphology and age from the other GC's in Fornax: 
it is a metal-poor
 cluster (${\rm [Fe/H]}=-2.01\pm 0.20$, on the Zinn \& West 1984 metallicity scale; 
 Buonanno et al. 1999) with a red HB and seems to be 3-4~Gyr younger than the other
clusters.
They also noted that the cluster CMD is virtually identical to that of the ``outer halo"
Galactic GC Ruprecht~106 (Rup~106), an unusual cluster often considered to have 
once been a member of a dwarf galaxy tidally disrupted by the Milky Way (e.g. Brown, Wallerstein,
\& Zucker 1997; Buonanno et al. 1999; Bellazzini, Ferraro, \& Ibata 2003).  

More recently, Strader et al. (2003) used low-resolution, integrated Keck spectra to measure metallicities 
and infer ages for the five GC's in Fornax. 
The new ages 
and metallicities are different than found by Buonanno et al. (1998, 1999).
%quoted in the introduction and used in the paper. 
According to Strader et al., the metallicity of For~4 is ${\rm [Fe/H]}=-1.5\pm 0.12$ (on the 
Zinn \& West 1984 metallicity scale),  
%not -2.0 as used in the paper 
and they estimate that the age of For~4 is the same as that of Fornax 1, 2, and 3, all older than 
Fornax~5 by 2-3~Gyr. 
%, contrary to statements 
%in the introduction. 

Because GC's are the oldest individual entities observable in galaxies, they provide invaluable 
insight into the formation history of the systems to which they belong.
A puzzling feature of the Galactic GC's is that they sharply divide into two distinct types according to
the mean periods of their RR Lyrae stars and the relative
proportions of fundamental-mode (RRab) and first-overtone pulsators
(RRc; Oosterhoff 1939). In the Milky Way (MW) 
%there are virtually no clusters filling the gap between 
Oosterhoff type I (OoI) clusters have $\langle P_{\rm ab} \rangle \simeq 0.55$~d, whereas Oosterhoff
type II (OoII) ones have $\langle P_{\rm ab} \rangle \simeq 0.66$~d (Clement et al. 2001). 
Galactic GC's belonging to different Oosterhoff types may have different
kinematical and spatial distributions,  possibly resulting from
different accretion/formation events in the halo (van den Bergh 1993;
Yoon \& Lee 2002, Miceli et al. 2007).
The existence of the Oo groups within the MW GC's can thus provide
clues about the halo formation process. However, we do not know
if the Oosterhoff dichotomy is a general characteristic of old
stellar populations in galaxies, or if it is
a peculiar phenomenon due to the particular evolutionary history of
the MW. Studying the Oosterhoff properties of GC's belonging to the dwarf companions
of the MW thus plays a crucial role in identifying the ``building blocks'' of the
Galactic halo and in understanding which galaxy formation scenario, whether 
the fast monolithic free-fall collapse of protoclouds
(Eggen, Lynden-Bell, \& Sandage 1962; but see also Sandage 1990) or 
the hierarchical merging and continual
accretion of lower-mass protogalactic fragments (Searle \& Zinn 1978), 
%
% (merger/accretion or
%loud collapse) 
is dominant (see, e.g., Catelan 2007).
%%{\bf Marcio, can you write the one or two sentences about these scenarios
%%that you suggest to add here? Thanks.}) 

Mackey \& Gilmore (2003a) identified candidate 
%presented a study of the 
RR Lyrae stars
in four of the Fornax dSph GC's (namely clusters \# 1, 2, 3, and 5) 
based on archival HST observations. 
They determined periods by fitting template RR Lyrae light curves to their data and 
%They conclude 
concluded that the Fornax clusters are unusual in
that their RR Lyrae populations have mean characteristics
 intermediate between the Oosterhoff groups.
% {\it (Qui va detto che per\`o i loro periodi sono fatti con curve
% template, ed infatti sbagliano completamente sulle RRd ad esempio.)}. 

For~4 has never been surveyed for variability, 
 in spite of clear indications, 
from its HB morphology (Buonanno et al. 1999), that it should contain RR Lyrae stars.
In this paper we present 
results from an extensive study of the variable star population in this extragalactic GC that 
resulted in the discovery of 29 variable stars, of which 27 are certainly \rrl\ stars.
The average luminosity of the  \rrl\ stars is used to measure the cluster distance. The 
%pulsation properties of the variable stars, namely 
periods, amplitudes, and period-amplitude distributions allow us to define the
Oosterhoff type (Oo-type) of For~4 and to verify whether it conforms to the Oosterhoff dichotomy shown by the 
Galactic GC's.

Observations and the adopted data reduction techniques are presented in Section~2. 
The variable star identification procedure is described 
in Section~3.
The pulsation properties of the For~4 variable stars and their membership in the cluster 
%and the Oosterhoff classification of For~4
are discussed in Section~4.
% and discussed in 
%along with the 
%comparison with 
In Section~5 we discuss the Oosterhoff classification 
and compare the For~4 Oo-type to that of 
%the Oo types 
%the Oo types of 
%of 
the other GC's in Fornax and to the MW GC's.  
%from Mackey \& Gilmore
%(2003) study,
% and to the Galactic GC's.
 In Section~6 we derive the cluster distance from the average luminosity of its
 RR Lyrae stars. Final results are summarized in Section~7.

%{\bf (Commento del referee: 
%
%The introduction is out of date in that it does not discuss 
%papers on Fornax globular clusters that are important in putting the current work in 
%perspective.  For example, 
%Strader et al. (2003, AJ, 125, 1291) used new Keck spectroscopy to measure metallicities 
%and infer ages for these five globular clusters; this work is not mentioned. The new ages 
%and metallicities contradict values quoted in the introduction and used in the paper. 
%According to Strader et al. the metallicity of For~4 is -1.5 not -2.0 as used in the paper 
%and the age estimate is that For~4 is same as 1-3, all older than 5, contrary to statements 
%in the introduction.  The authors don't have to agree with Strader et al. but they should 
%address the differences. ---- 
%
%Risposta di Marcio: 
%
%adopting a metallicity of 
%[Fe/H] = $-$1.5 according to Strader et al. (2003, AJ, 125, 1291), would place Fornax 
%4 right in the middle of the Oosterhoff-
%intermediate band in the [Fe/H]-HB type diagram (Fig. 8 in my Canc\'un
%paper), whereas our currently adopted [Fe/H] ~ $-$2 actually places it a
%bit to the red of this band. None of the conclusions based on P$_ab,min$
%are affected in any way by a change in the metallicity.}

\section{Observations and data reductions}
%%\subsection{Observations}

$B,V$ time series photometry of For~4 and of its surrounding field was obtained with the MagIC camera of the 
Magellan/Clay 6.5-m telescope on two nights in November 2003 and on two nights in December 2004.

%---------------------------------------------------------------
We covered a total field of view of $2.4^{\prime} \times 2.4^{\prime}$,
centered on the cluster at RA=2:40:07.33, DEC=$-$34:32:17.4\ (J2000).
Nights were photometric with seeing conditions varying from  $0.45\arcsec$ to  $0.65\arcsec$.
%$^_{\prime}$$^_{\prime}$ and $0.65^_{\prime}$$^_{\prime}$.
%Our observations consist of 4 nights over one year, from November 2003 to December 2004. The scheduling was carefully 
%planned in order to get a good coverage of the light curves specifically for short-period - like \rrl- variable stars.
We acquired a total number of 58 $V$ and 19 $B$ images. Average exposure times were 
500 seconds for the $V$ frames and 700s for the $B$ ones.
Observations of the standard fields T~Phe and Ru~149 (Landolt 1992) 
were obtained on the same nights, to calibrate the 
data to the standard Johnson-Cousins photometric system.
	    
The Magellan data were complemented by observations of the Fornax dSph field containing 
For~4\  obtained with the wide field imager of the 4-m Blanco/CTIO telescope. Data  
%For~4\ was imaged on CCD n. 1 of the Blanco mosaic.
%
%Data from the 4-m Blanco Telescope 
were collected over three runs for a total of 5 nights spread over three years, from 2003 to 2005.
%covering a period of three years, for a total of 5 nights  (Table 1). 
 The images were taken with 
 the Mosaic II imager, which has a $36^{\prime} \times 36^{\prime}$ field of view centered at
 RA=2:41:07.29 and DEC=$-$34:16:35.6 (J2000) with a plate-scale of $0.27\arcsec/{\rm pixel}$.  
 The cluster was imaged on amplifier 1 of the 16-amplifier
 mosaic.  The conditions during the three runs were photometric or
 nearly photometric.  The seeing varied over the range of $0.8\arcsec$ to $1.9\arcsec$.
 The exposure times varied from 100s to 600s in $V$ and 200s to 600s in
 $B$ with 100s and 200s being typical, respectively.
%%%{\bf Horace, Armin and Nathan, could you please provide details for the CTIO observations}
The CTIO dataset consists of 142 $V$ and 61 $B$ images.
%%%
Logs of the observations and details of the instrumental set-up at the two telescopes are given in
Table~1. 

The Magellan frames were bias-subtracted and flatfield-corrected using the MagIC tool in 
IRAF.\footnote{IRAF is distributed by the National Optical Astronomical 
Observatories, which are operated by the Association of Universities for
Research in Astronomy, Inc., under cooperative agreement with the 
National Science Foundation.} They were then reduced with DAOPHOT-ALLSTAR.
Time-series photometry for all the stars was produced with ALLFRAME (Stetson 1994, 1996).
The photometric precision at the HB level is 0.01~mag.
The photometric calibration was derived using the standard stars
observed at the Magellan telescope on the night of Nov. 13, 2003 (UT).  

A set of linear calibration
relations was computed: 
\begin{center}
\begin{eqnarray*}
B  &=&  b\arcmin  +  0.051 (\bv)  +  26.87, \\
V  &=&  v\arcmin  -0.049 (\bv)  + 26.81, 
\end{eqnarray*}
\end{center}
\noindent 
where $b\arcmin$ and $v\arcmin$ are the instrumental magnitudes
normalized to 1-s exposure and corrected for atmospheric extinction by
adopting the mean extinction coefficients $k_B = 0.22$ and $k_V = 0.12$.
The zero point uncertainties of the calibration relations are estimated
to be of the order of 0.05~mag in $B$ and $V$, and 0.03~mag in $(\bv)$,
from the
night-to-night scatter of the standard stars.
These relations were used to calibrate the instrumental magnitudes,  
%in our photometric catalogs 
%along with 
and aperture corrections 
%to align the 
%point spread function ({\sc psf}) photometry, that were  
%estimated by performing  
%aperture photometry on 
were applied based on photometry of relatively isolated stars.

%%%%%%%%%\realfigure{f1.eps}
%%%%%%%%%{Color-magnitude diagram of For~4\ from the Magellan data. 
%%%%%%%%%Variable stars are displayed with different symbols.
%%%%%%%%%Open circles: {\it ab-}type RR Lyrae stars; asterisks: first overtone RR Lyrae stars;
%%%%%%%%%filled circle: double-mode pulsator (RRd). The open triangle marks a peculiar 
%%%%%%%%%RR Lyrae star (V13).}{f:fig2}

The Magellan observations of For~4\ produced a CMD reaching $V \simeq 25$~mag. 
This CMD is shown in Figure~1. Although the cluster area observed with the Magellan telescope mainly 
contains stars within the For~4\ tidal radius, the cluster CMD is clearly 
contaminated by stars belonging to the Fornax field, since the cluster 
lies in the center of the Fornax dSph galaxy. 
%This is clearly seen in Figure~1. 
On the other hand, no reliable photometry could be obtained for stars  inside a 
radius of $7 \arcsec$ from the cluster center, due to crowding. 
The CMD in Figure~1 was drawn cutting off this central region.

The images taken at the 4-m Blanco telescope were reduced using the SMSN
pipeline (Rest et al. 2005) originally created for the SuperMACHO and
ESSENCE projects. The SMSN pipeline reduces the raw images using the
standard IRAF mosaic reduction package, MSCRED, as well as its own routines.
As part of this reduction, the images were cross-talk corrected,
astrometrically calibrated, and split into the 16 separate amplifiers.  The
cross-talk correction is necessary to remove ghost images of bright stars
that can appear as a result of the electronics of the mosaic imager.  Since
the images are optically distorted due to the large field of view, they were
then de-projected into a tangential plane using a customized version of
Swarp (Bertin et al. 2002).
%
%The images taken at the 4-m Blanco telescope were reduced using the SMSN
% pipeline (Rest et al. 2005) originally created for the SuperMACHO and ESSENCE
% projects.  The images were cross-talk corrected
% 
% ({\bf Punto 5 dei commenti di Marcio, NATHAN}), astrometrically
% calibrated, split into the 16 amplifiers, and standard reduced. Since the
% images are optically distorted due to the large field of view, they were
% then deprojected into a tangential plane using a customized version of
% Swarp (Bertin et al. 2002).  
 DoPHOT (Schechter, Mateo, \& Saha 2003) was then used on the frames to
 get profile-fitted photometry of the stars (and other objects),
 with photometric precision 
 of 0.03-0.05~mag for stars on the HB.
 Local standards were used to 
 transform individual images to the standard system. Final 
 photometric calibration was then obtained by adjusting 
 to the zero points of the Magellan dataset. 

\section{Variable star identification}
%%\subsection{Magellan data}

Variable stars in For~4\ were identified by applying the Image Subtraction Technique, as 
performed by ISIS2.1 (Alard 2000), to the 2003+2004 Magellan $V$ dataset (58 frames in total),
producing a catalog of about 400 candidate variable stars. 
The ISIS catalog was  cross-identified against the DAOPHOT/ALLFRAME photometric catalog, 
leading to light curves in magnitude scale and confirming the variability for 20 stars.
A further 9 variable stars were recovered in the cluster's central region where we lack reliable 
{\sc psf} photometry. For these stars we analyzed only the differential
flux light curves produced by ISIS. 
%%%Further 9 variable stars were recovered in the cluster central region were we
%%%lack reliable photometry. 
%%%For these stars we analyzed the differential flux light curves. 

 Variable stars were independently identified from the CTIO dataset, using difference image analysis 
 (Phillips \& Davis 1995; Alard \& Lupton 1998; Alard 2000). 
 We used the HOTPANTS package
 (Becker et al. 2004) to match the {\sc psf} of the image and
 template by applying a spatially varying kernel to the image with the
 better seeing before the template is subtracted. 
 %%%%%%% ACCORCIATO
One of the main problems
 with the image differencing approach is that there are more residuals,
 e.g. cosmic rays and bleeds, than genuinely variable objects in the
 difference image. Therefore standard profile-fitting software like
 DoPHOT has problems determining the proper {\sc psf} used to perform photometry
 in the difference image. %When the 
 The difference image was analyzed with our
 customized version of DoPHOT, where we forced the {\sc psf} to be the one determined
 for the original, flattened image.  Applying this a priori knowledge of
 the {\sc psf} helped to guard against bright false positives, such as cosmic rays
 and noise peaks, which generally do not have a stellar {\sc psf}.
%%%{\bf Armin and Nathan: could you please describe the procedures for the photometric reduction and 
%%%the identification of the variable stars adopted 
%%%for the CTIO dataset}

By matching the coordinates we were able to cross-identify 20 of the variable stars
identified from the Magellan dataset. Fifteen of them had been 
independently recognized as variable stars from the CTIO data.
Due to the high crowding no reliable match could be obtained for variable stars
in the cluster central region.
The time series $B,V$ photometry of the 20 confirmed variable stars with light curves
in magnitude scale is provided in Table~2 and is available in electronic form 
in the electronic edition of the journal. 

%%\subsection{Variable star catalogues}
%%
%%{\it Qui ci va la presentazione delle tabelle dati. Inserire tabella dati:
%%3177.tex.}

\section[]{Period Search and Pulsation Properties of the Variable Stars} 

%%%%The possibility of combining the two datasets allowed
%%%%us to increase the time--on--targets, greatly improving
%%%%the phase coverage. In such a way we could solve the
%%%%alias ambiguities at $f~\pm~1~$cd$^{-1}$, which were
%%%%very severe when considering only one dataset. The
%%%%datasets merging also returned a better estimate of 
%%%%the light curve parameters.

%%%%Both  the Magellan and CTIO data-sets  
%%%%were used to determine the pulsation characteristics of the variable stars 
%%%%from the study of the light curves.
%%%%The possibility of combining such a large amount of observations 
%%%%%over such a long time baseline 
%%%%helped to solve ambiguities in the period determinations,  allowing at the same time a good 
%%%%estimate of the star pulsation parameters.

The time series of the 400 candidate variable stars were analyzed 
%Periods, epochs of maximum light and amplitudes of the light variation of the variable stars were derived 
independently by two of us (GC and EP) using two 
different codes and then compared. GC used  
GrATiS (Graphical Analyzer of Time Series), a custom 
package developed at the Bologna Observatory by P. Montegriffo (see Di Fabrizio 1999 
and Clementini et al. 2000), which uses 
both the Lomb periodogram (Lomb 1976, Scargle 1982) and the best fit of the data with
a truncated Fourier series (Barning 1963).

EP used the iterative sine--wave least--squares
 method (Vanicek 1971), combined with a Fourier
 series decomposition limited to significant 
 terms only (Poretti 2001). Visual inspection of
 the intranight light curves was also performed. 
% 
%({\bf Punto 6 di Marcio, mandato mail ad Ennio, direi pero' che
%qualunque variazione che si riteneva vera sia che producesse o meno un P definito 
%l'abbiamo tenuta in conto, quindi direi che si puo' escludere che abbiamo cacciato via
%possibili candidate variabili di lungo periodo o irregolari.
%Ad esempio in Chip6 troviamo una long period quindi la tecnica non e' biased 
%verso le irregular o le long period.}) 
%
After having compared the results, only candidate variable 
 stars confirmed by both methods have been accepted (29 stars in total).
Our time series extends over three consecutive nights in 2003, 4 
nights in 2004, and 1 further night in 2005 (see Table~1). 
Variable stars with periods of several days or longer would not be detectable in
the data set of each individual year, since the long period will show effects only on longer timescales. 
%
%they would not appear to vary on the few
%nights of each year
%(the long period will  show effects only on a longer timescale), but 
Such long-period variable stars, if present, would manifest themselves as stars having different
magnitudes in the three years of our survey; since we have only 4-5 epochs, they would appear as periodic
variable stars  with a prominent low-frequency peak in the power spectra. Indeed, we do not 
have such candidates in our sample.
At the same extent, non-periodic variable stars with timescale of 1-5 days
will result in periodic variable stars, since any periodogram will find a (false) period
when a non-periodic variable is sampled at 4-5 epochs only.
% a
%they have been observed on four-five nights only.
%In a certain sense, when you sample a non-periodic variable at 4 epochs
%only,
%
Irregular, rapid variable stars are of course undetectable, but they are also not
very probable. In conclusion, it is rather unlikely that we have missed any such
long-period variable stars in For~4, if any is present.

Both  the Magellan and CTIO datasets  
were used, when available, to determine the pulsation characteristics of the variable stars 
 from the study of their light curves. The possibility of combining the two datasets allowed
us to increase the time--on--targets, greatly improving
the phase coverage. In such a way we could solve the
alias ambiguities at $f\pm 1$~cycle/day, which were
very severe when considering only one dataset. The merged
datasets also returned a better estimate of
the period, epoch of maximum light and amplitudes of the light variation of the 
confirmed variable stars.
%the light curve parameters. 
 We were able to derive reliable periods and light
curves for 20 variable stars observed with both the 
%namely all the variable stars with data from both the 
Magellan and CTIO telescopes and for 3 stars with differential fluxes from the Magellan dataset only.
%When working with just one dataset the poor phase 
%coverage and the aliasing problems affect the 
%period determination. Thus, on 
For the remaining six cases 
we can only provide a rough estimate of the period or
simply an identification of variability. 
%%%Star V5 constitutes a particular case: 
The spectral window of our data is dominated by strong
aliases at $n$ cycles per year (with $-3\le n \le 3$). 
We used period values to the fifth decimal place to fold the light
curves, since this allowed us to significantly
reduce the r.m.s. scatter 
%of the light curve best fitting models. 
 of the truncated Fourier series best fitting the data.
However, our period determinations are not free from 
the annual aliasing effect, and this can easily affect the fourth
decimal place. In other words, the true frequency could be an alias
at $n$~cycle/year of that corresponding to the periods finally adopted 
for the variable stars (see Table~3).
The standard deviations of the least--square fits of the Magellan light curves are
in the range from 0.02 to 0.04~mag and from 0.01 to 0.07~mag in $V$ and $B$, respectively---which is 
quite acceptable and comparable with the expected precision of stars with $V\sim$~21.5~mag in a very 
crowded field.

Twenty-seven of the variable stars detected in For~4 clearly appear to be RR Lyrae stars: 
22 {\it ab-}, 3 {\it c-}type, and 2 double-mode pulsators. However, among the listed fundamental-mode pulsators three have 
uncertain classifications, and 
a fourth one, star V13, has the typical period of a fundamental-mode pulsator, but the light curve shape of 
a first overtone variable star. 
 Although the star shows significant gaps in the  light curve, we could not fit the data with a shorter period.
While long-period RRc stars have been observed in a few GC's 
(namely $\omega$~Centauri~= NGC~5139, NGC~6388, NGC~6441, and one star 
 in M3; see Catelan 2004b), such stars tend to be bright, whereas V13
  appears to be even fainter than average, which is quite unusual 
  and makes it very hard to explain its long period if it is indeed
  a first-overtone pulsator. Also, if ${\rm [Fe/H]}\sim -2$, For~4 would by far
  be the lowest-metallicity globular cluster known to harbor
  a long-period RRc. Such circumstantial evidence may be in favor of
  our RRab, as opposed to a long-period RRc, classification for V13 
  (see Table~3). We note that 
   an RRab star with behavior similar to V13 was observed in   
   NGC~6441 by Pritzl et al. (2003).

Coordinates for the variable stars are 
provided in Table~3, along with period, type, time of maximum light, number of phase points in the
Magellan and CTIO datasets separately, intensity-averaged $\langle V\rangle$ and
$\langle B\rangle$ magnitudes, and amplitudes of the light variation ($A_V$ and $A_B$).  
%%{\bf [Should we provide magnitude-averaged magnitudes/colors as well? Also, should we 
%%state whether we applied the amplitude-dependent correction to the ``equivalent static 
%%star'' to produce Fig.~1 (from Bono et al. 1995), and if not, explain why? --MC]}
%
%available form the authors. 
Examples of light curves are shown in Figure~2. 
%%{\bf [I think it's not really necessary (i.e., it's redundant) to present this figure, 
%%since it does not seem to add anything relevant to what is included in the Appendix. --MC]} 
 The complete atlas of light curves is presented in the Appendix.
%
%
%complete catalog of light curves is published in the electronic edition of
%the journal.}
 %Of the stars listed in Table~3, 
Variable stars V8 and V9 have lights curve only in differential flux and large  
gaps around the maximum light (see Fig.~5). In turn, their period and classification
in type are very uncertain.
The folded light curves of V14 
show scatter larger than for the other variable stars, in the CTIO dataset
in particular (see Fig.~6), which renders the star's
mean magnitudes and amplitudes uncertain. However, the individual measurements
seem to confirm the intrinsic faintness of V14, which   
appears to be about 0.6~mag fainter than the average level of 
the cluster's HB. 
%For sake of completeness and despite of the uncertainties, the star is reported
%in Fig. 1with an errorbar of $\sim 0.3$~mag.
This faint magnitude would put V14 likely  
%does not belong to For~4. 
beyond the limits of Fornax if the star is a {\it bona-fide} RR Lyrae variable.
%%%Could this variable be a high amplitude $\delta$ Scuti star? In effetti e' una stella 
%%%molto blu (la seconda piu' blu dopo la V13 che e' quella con curva da c e periodo
%%%da ab), ed ha ampiezze sia $V$ che $B$ molto grandi, non potrebbe essere qualcosa
%%%di diverso invece che una RR? discuterne con Ennio.
Stars V10, V15 and V29 also are interesting given their 
small amplitudes for their periods. 
The Magellan light curves of V5 clearly show maxima 
at different heights, suggesting a double-mode nature, however we could not 
derive reliable periodicities for the star.
%%%Period-amplitude distributions for the For~4 RR Lyrae stars 
%%%with light curves in magnitude scale are shown in Figure~3, where the periods of the 
%%%{\it c-}type RR Lyrae stars 
%%%have been transformed to fundamental mode periods by assuming log~P$_f$ = log P$_c$ +0.128. 

\subsection{Membership Probability}

%{\it (Comment del referee:
%
%Membership probability  FATTO USANDO LE STIME DAI NOSTRI DATI CTIO COME RICHIESTO 
%                        DAL REFEREE
%
%The authors use a reasonable method to infer the contamination of their 
%globular cluster RR Lyr sample with field RR Lyr stars if all they have 
%is a small field.  But the CTIO data they use encompass a field of 36x36 
%arcmin, much larger than For~4.  Why did they not use this field to determine 
%the field RR Lyr density for use in defining the surface density of RR Lyr stars?  
%Secondly, Bersier and Wood (2002, AJ, 123, 840) report detection of some 
%500 RR Lyr in the Fornax dSph galaxy.  This work must surely impact their 
%knowledge of the field density of RR Lyr in Fornax. 
%Thirdly, these authors participated in a recent study of the variable 
%star population in the Fornax field that found ~1000 candidate variables 
%(Maio et al. astro-ph/0303666) which result could well have been used in 
%the current work.)}  
%
%

Given the different stellar content of For~4 and the
surrounding Fornax field populations, estimating the expected number of
RR\,Lyrae variable stars belonging to For~4 from the overall surface density profile is not
straightforward (see Mackey \& Gilmore 2003b). 
We have used both the outermost variable stars in the Magellan
field and the variable stars found in the larger CTIO field to empirically
estimate the expected surface density of RR\,Lyrae in the Fornax galaxy.
In the Magellan data, 6 variable stars have been identified beyond
50\arcsec, in an area of $1.23 \times 10^{4}\,{\rm arcsec}^2$.  On the basis
of the overall surface density profile of For~4, we can assume that these
variable stars are not associated with the cluster.
Then we estimate a surface density of $4.9 \times 10^{-4}$ RRL per
arcsec$^2$ in the Fornax field. As for the CTIO data, RR Lyrae stars have
been counted in an annulus between 140\arcsec\ and 220\arcsec\ from the
center of For~4, which is an optimal compromise between the statistical
benefits of sampling a large area and the need to keep our measurement
{\it local} in view of the spatially varying stellar populations of
Fornax.  A surface density of $7.2 \times 10^{-4}$ RRL~arcsec$^{-2}$ is
inferred from the CTIO dataset.  In the inner region, within a radius of
$30\arcsec$ from the center, 1.4 and 2.0 field RR Lyrae stars are
therefore expected, respectively.
In conclusion, among the RR Lyrae stars within about $30\arcsec$ from
Fornax~4, about 2 stars possibly belong to the Fornax dSph field.

The same exercise for the innermost variable stars found with ISIS shows that
they are most likely to be all cluster members.

%%\section[]{Discussion} 
%%{\it Ho inserito brutalmente la nota di Enrico sulla contaminazione statistica}
%%\subsection{Radial profile}
%%%%%%%%%%%%%%%%%%%%%%%%%%%%%%%%%%%%%%% tex di enrico
%%Figure~\ref{fig:surfdens} shows the surface density of stars around
%%FGC4. The count excess produced by the stars in the globular cluster
%%does not extend beyond $R=50\arcsec$. This is consistent with the
%%density profile of FGC4 obtained by \citet{mack+gilm03}.  See the SM
%%procedure {\sc diagramma.sm} for details.
%%
%______________________________________________
%%\begin{figure}
%%
%%\caption{Surface density of stars in the Fornax globular cluster 4 and
%%surrounding field.  Stars have been counted in circular annuli with
%%2\arcsec\ width, divided by the area of the annulus, and plotted against
%%radial distance from the center of FGC4, assumed to be at pixel
%%1201,885. No completeness correction has been applied. We also plot
%%error bars according to Poisson statistics. The continuous line is a
%%visual fit using an EFF profile with the $a$,$\gamma$ parameters given
%%by Mackey \& Gilmore (2003), and the amplitude fitted by eye.  A constant
%%was added to the EFF profile to fit the contribution of the Fornax
%%field.  This profile, which essentially reflects the distribution of RGB
%%stars, confirms that GC4 extends out to $\sim 50\arcsec$.}
%%\label{fig:surfdens}%
%%\end{figure}
%---------------------------------
%\subsection{Average periods and Oosterhoff classification}
%
\section{Oosterhoff type and metallicity}
% of the For~4 RR Lyrae stars} 
%\section{Discussion}

%%{\it (Comment del referee:
%%
%%Discussion
%%
%%With the new spectroscopic metallicity for For~4, which is likely to be much 
%%more reliable than the Buonanno et al value (1999), the metallicity of For~4 
%%is near -1.5 rather than -2.0.  How does the discussion of the Oo group 
%%(via Sandage's method) change if metallicity is -1.50?
%%This is the only section of the paper that I found confusing.  I think it 
%%would help the presentation of their results if they put the Oosterhoff 
%%numerical comparisons into a table rather than have it all in the text.
%%There is no comparison of their distance value with other values in the 
%%literature for Fornax; there should be.  With the greater metallicity their 
%%distance to For~4 changes to 20.69 mag. The authors note that their mean RR 
%%Lyr V mag agrees extremely well with the mean HB V value from Saviane et al 
%%(2000, A\&A, 355,56); but they don't mention that Saviane et al infer a distance 
%%modulus from the HB of 20.76+/-0.04 which is markedly different from the result 
%%in this paper prior to changing [Fe/H] to -1.5.  Similarly, Mackey and Gilmore 
%%find distance moduli for the other four globular clusters in Fornax to be 20.58 - 20.74. 
%%With the revised metallicity the authors' distance estimate is in much better agreement
%% with other published values.)}
The pulsation properties of the For~4 RR Lyrae stars are summarized in Table~4, 
where we list the  
number of fundamental-mode (RRab), first overtone (RRc) 
and double-mode pulsators (RRd) in the first three columns, respectively; 
the ratio of number of RRc to total
number of RR Lyrae stars (RR$_{tot}$), in column~4; the average periods of the {\it ab-} and 
{\it c-}type RR Lyrae stars, in columns~5 and 6; and the shortest and longest RRab periods,
in columns~7 and 8. 
%%{\bf [Why do you give the longest RRab period? --MC]} 
The average period of the {\it c-}type RR Lyrae stars is
$\langle P_{c}\rangle=0.360$~d ($\sigma=0.042$~d, 
average on 3 stars), or 0.374~d ($\sigma=0.044$~d, average on 4 stars), if the RRd star is included. 
The {\it ab-}type RR Lyrae stars have 
%{\it ab-}type RR Lyrae stars is: 
$\langle P_{ab}\rangle=0.594$~d ($\sigma=0.054$~d, average on 
19 stars). The quoted sigma values are the dispersion of the mean. 
V13 and V14 were not considered. Average and $\sigma$ do not change
if V13 is included, but $\langle P_{ab}\rangle=0.596$~d if V14 is added.
The shortest-period RRab is star V16 with $P_{ab,min}=0.5191$~d, and the
longest-period RRab is star V29 with $P_{ab,max}=0.6986$~d.

%{\bf [Why do you start the preceding paragraphy with ``ab, c, d'' in ``RRab,c,d'' as subscripts, 
%only to finish it going back to the more usual, non-subscript form? We should probably 
%adopt a single form throughout the text. --MC]} 
%
%
%
%The {\it c-}type RR Lyrae stars have 
%for the {\it ab-} and the {\it c-}type RR Lyrae stars, respectively.
The period distribution of the RR Lyrae stars in For~4 is shown in Figure~3. The right panel 
shows basically the same distribution as in the left panel, only with the RRc periods ``fundamentalized'' 
according to the relation $\log P_{f} = \log P_{c} + 0.128$. The
RRab distribution may give some hints of having two separate peaks at
$\langle P_{ab}\rangle=0.55$~d and $\langle P_{ab}\rangle=0.64$~d, respectively---thus
resembling the case of NGC~1835 in the Large Magellanic Cloud (Soszynski et al. 2003). 
%%{\bf [I'm afraid this is not really evident from the histogram!? --MC]} 
However, a KMM test (Ashman, Bird, \& Zepf 1994) performed on our data shows that 
they  are consistent with a unimodal distribution in $P_{ab}$. Interestingly, the 
fundamentalized period distribution is peaked at a value $\approx 0.6$~d, which is 
significantly longer than for either OoI or OoII clusters. We refer the reader to 
Catelan (2004a) for an extensive discussion of the implication of the peaked 
distributions in fundamentalized RR Lyrae periods.

According to the RR Lyrae star average periods For~4 appears to be an 
Oosterhoff-intermediate (Oo-Int)
cluster, an Oo-type that has very few counterparts among the Milky Way GC's 
(Catelan 2004b, 2005). 
% with a confidence level of 
%the probability
%of having two separate populations is less than 18\%. 
%82\%.
%({\bf Fare tutti i vari confronti delle Oosterhoff numerical comparisons in una tabella.)} 
Even if we only consider stars in the inner region of the cluster ($R \le 30\arcsec$) 
and if we also take into account that one or two of them could
not belong to the cluster (see Section 4.1), we have $\langle P_{ab}\rangle=0.594$~d, thus 
confirming a true Oo-Int behavior, similarly to what was found for
the other 4 GC's in Fornax by Mackey \& Gilmore (2003a).

Period-amplitude distributions for the For~4 RR Lyrae stars 
with light curves in magnitude units are shown in Figure~4. 
%%%where the periods of the 
%%%{\it c-}type RR Lyrae stars 
%%%have been transformed to fundamental mode periods by assuming log~P$_f$ = log P$_c$ +0.128.
Expanded symbols are used for stars located at distances less than 
$30\arcsec$ from the cluster center.  
Figure~4 shows that the majority of the  {\it ab-}type RR Lyrae stars in For~4
appears to be on the OoI line. It is also noteworthy that the ratio of RRc's to the total
number of RR Lyrae stars is more like OoI (see Table~4).  
%(RRc/RRtot = 0.12 without 
%the RRd stars and 0.19 with them, or  0.16 and 0.20, respectively, 
%depending upon whether we consider all the RR Lyrae
%stars in For~4, or just those within $30\arcsec$ from the cluster center).
In Figure~4 there are, however, a number of RRab's that pile 
up near the OoII location.
%, as if two separate populations were present within the
%cluster.  Moreover, 
%%%%%%%%%\realfigure{f3.eps}{$V$, $B$ period-amplitude diagrams of For~4 \rrl\ stars.
%%%%%%%%%% with light curves in magnitude scale.  
%%%%%%%%%Symbols are as in Figure~1, but those indicating variable stars located at distances
%%%%%%%%%larger than $30\arcsec$ from the cluster center are expanded.
%%%%%%%%%%Filled and open circles are {\emph c-}
%%%%%%%%%%and {\emph ab-type} \rrl\ stars, respectively. The x sign is V13.
%%%%%%%%% The straight 
%%%%%%%%%%dashed 
%%%%%%%%%lines
%%%%%%%%%are the positions of the OoI and OoII Galactic GC's according to Clement \& Rowe (2000). 
%%%%%%%%%Period-amplitude distributions of the
%%%%%%%%%{\it bona fide} regular (solid curves) and well-evolved (dashed curves) {\emph ab} \rrl\ stars in M3 
%%%%%%%%%from Cacciari, Corwin, \& Carney (2005) are also shown for comparison.}{f:fig3}
However, the two stars closer to the OoII line are located
outside $R=30\arcsec$, thus their membership is less certain.
%may belong to the field.
%It is also noteworthy that the ratio of RRc to the total
%number of RR Lyrae stars is more like OoI (RRc/RRtot = 0.12 without the RRd stars and 0.19 with them).
%%%According to the RR Lyrae star average periods For~4 appears to be an 
%%%Oosterhoff-intermediate (Oo-Int)
%%%cluster, an Oo-type that has no counterpart among the Milky Way GC's.

As discussed in the Introduction, the metal abundance of For~4 is somewhat 
uncertain, being ${\rm [Fe/H]}=-2.0$ in the Buonanno et al. (1999) study, and
${\rm [Fe/H]}=-1.5$ in Strader et al. (2003).
There are some metallicity estimates that can be made using the 
pulsation properties (namely periods and amplitudes) and the shape of the light curve
(namely the parameters of the Fourier decomposition of the light curve) of the RR Lyrae stars. 
%I will leave it up to you if 
%you would want to include them.  
Applying the Alcock et al. (2000) relation, which is based on the period and $V$-band
 amplitude of RRab stars, we found a mean metallicity of $-1.68$ for the cluster
 (or $-1.66$ for the 10 RRab stars within 30\arcsec\ from the cluster center). 
 This is intermediate between the Buonanno et al. and Strader et al. values.
%%slightly more
%%metal-rich than found by Buonanno et al. (1999).  
Using relations from Sandage (2006), we find [Fe/H] values
of $-1.96$ from the mean period of the fundamental-mode RR Lyrae stars, 
% {\rm $\langle P_{ab}\rangle$}, 
$-2.25$ from the shortest-period RRab (V16, a star located within $30\arcsec$ from the 
cluster center, hence very likely belonging to For~4), and $-1.38$ 
from the longest-period RRab (star V29, the most external of the variable stars
in our sample, hence probably belonging to the Fornax dSph field). 
%%{\bf [The referee may 
%%wonder why we are doing this for individual cluster stars, since a scatter in [Fe/H] may 
%%be implied by these results, if taken at face value. --MC]} 
Finally, from Sandage (1993), we find ${\rm [Fe/H]}=-1.90$ from the mean period of the three
cluster RRc stars, 
%{\rm $\langle Pc\rangle$}, 
which are all located within $30\arcsec$ from the For~4 center.

According to Jurcsik \& Kov\'acs (1996) the [Fe/H] abundance of a fundamental-mode 
RR Lyrae star is a linear function of the period $P$ and of the  
parameter $\phi_{31} \equiv \phi_3 - 3\phi_1$ of the Fourier decomposition of the star's $V$-band light curve. 
Individual metal abundances were derived for the For~4 fundamental mode RR Lyrae stars
%from the Fourier parameters
%parameters of the Fourier 
%decomposition 
%of the $V$ light curves 
%of the fundamental mode RR Lyrae stars 
by applying the 
Jurcsik \& Kov\'acs method (equation~3 in their paper) 
%{\it photometric} metallicities for our
to the RRab variable stars 
whose light curves satisfy  the {\it compatibility conditions} (namely 
light curve completeness and regularity criteria) 
%%, referred to by these
%%authors as {\it compatibility conditions}, 
for the Fourier parameters to predict reliable empirical quantities
(see Kov\'acs \& Kanbur 1998).\footnote{
According to Kov\'acs \& Kanbur (1998), 
deviations of the Fourier parameters  
should not exceed the 
maximum value ($D_{\rm m}$) of 3, with 
maximum deviations  $D_{\rm m} > 3$ possibly 
indicating incompatibility with the empirical predictions.
The deviation parameters $D_{F}$ are defined as     
$D_{F} = \mid F_{obs} - F_{calc} \mid / \sigma_F$, where  
$F_{obs}$, $F_{calc}$ are respectively the observed value of a given
Fourier parameter and its predicted value from the other observed
parameters, and  $\sigma_F$  is the respective standard deviation 
(see eq.~6  and Table~6 of Jurcsik \& Kov\'acs 1996).} 
% whose light curves satisfy completeness and regularity criteria
%%, referred to by these
%%authors as {\it compatibility conditions}, 
%%for the Fourier parameters to predict reliable empirical quantities
%%(see Kov\'acs \& Kanbur 1998, hereinafter KK98).   
%XXX of the For~4 fundamental mode pulsators satisfy
%
%subsample of 29 {\it ab-}type RR Lyrae stars
%using 
%equation (3) of JK96 (see also K02).
%; hereinafter JK96),
%, A\&A, 312, 111), 
%Kov\'acs \& Jurcsik (1996, 1997; hereinafter KJ96, KJ97),
% ApJ, 466, L17; 1997,  A\&A, 322, 218) 
%and   
%Kov\'acs  \& Walker (2001; hereinafter KW01) relations
%to RRab variables whose light curves satisfy completeness and regularity criteria
%%, referred to by these
%%authors as {\it compatibility conditions}, 
%for the Fourier parameters to predict reliable empirical quantities
%(see Kov\'acs \& Kanbur 1998, hereinafter KK98).  
%% Namely, the deviations of the Fourier parameters  
%% should not exceed the 
%% maximum value (D$_{\rm m}$) of 3, with 
%% maximum deviations  D$_{\rm m} > 3$ possibly 
%% indicating that incompatibility with the empirical predictions can 
%% be expected (Kov\'acs \& Kanbur 1998, hereinafter KK98).
%% The deviation parameters D$_{F}$ are defined as     
%% D$_{F}$ =$\mid F_{obs} - F_{calc} \mid / \sigma_F$, where  
%% $F_{obs}$, $F_{calc}$ are respectively the observed value of a given
%% Fourier parameter and its predicted value from the other observed
%% parameters, and  $\sigma_F$  is the respective standard deviation 
%% (see eq. 6  and Table 6 of JK96). 
Two of the For~4 RRab stars are found to satisfy these conditions, namely star V18 and V24, 
%{\it compatibility conditions}.
for which we derive ${\rm [Fe/H]}=-1.62$ and $-1.82$, respectively.
These {\it photometric} metallicities
are on the Jurcsik (1995) 
metallicity scale and can be transformed to the Zinn \& West (1984) scale
by applying the transformation relation provided by equation~4 in Jurcsik (1995),
leading to ${\rm [Fe/H]}=-1.75$ and $-1.89$, respectively. 

%
%Parameters from the Fourier decomposition of the $V$ light curves of these
%stars  
%are provided in Table~4  along with the respective highest D$_{\rm m}$ values.
%The derived metallicities are listed in Column 4 of Table~5, their
%uncertainties were calculated 
%from eq.s (4) and (5) of Jurcsik \& Kov\'acs (1996), and
%adopting for the Fourier parameters the standard deviations provided in 
%Table~2 of Kov\'acs \& Kanbur (1998).
%%along with the related uncertainties that 
%%were evaluated 
%%%Errors were calculated 
%%%from eq.s (4) and (5) of Jurcsik \& Kov\'acs (1996), and
%%%adopting for the Fourier parameters the standard deviations provided in 
%%%Table~2 of Kov\'acs \& Kanbur (1998). 
%%
%Metallicities ([Fe/H]) are provided in Column 4 of Table, 
%These {\it photometric} metallicities 
%are based on Jurcsik (1995) 
%metallicity scale. They 
%span the range: $-0.XX<$[Fe/H]$<-1.XXXX$, with   
%average value [Fe/H]=$-1.XXX$ ($\sigma= 0.XX$, XX stars), and mean uncertainty of 
%about 0.XXX dex (see column 4 of Table~5)
%for stars closer to the cluster center and XXX XXXX for the remaining 
%stars XXXXX, corresponding to XXXX and XXX respectively 
%in the Zinn \& West (1984) metallicity scale.
Thus, different methods based on the cluster RR Lyrae stars 
%that more likely are cluster members 
seem to suggest a low
metal abundance for For~4, closer to the Buonanno et al. (1999) estimate.
However, we should be aware that the calibrations of both the 
period-amplitude-[Fe/H] ($P-A-{\rm [Fe/H]}$) and the period-$\phi_{31}$-[Fe/H] ($P-\phi_{31}-{\rm [Fe/H]}$) relations
%
%the relation and calibrations and the 
are based on MW GC's that do follow the Oosterhoff dichotomy,
and might thus not be suitable to derive the metal abundance 
of an object that does {\em not} follow the dichotomy such as For~4.
 Also, the $P-A-{\rm [Fe/H]}$ relations have a known 
 % notoriously
 % known for the amount of 
 scatter 
 %they present 
 at any given [Fe/H]. For
  instance, there is a range of [Fe/H] where OoI and OoII objects
  overlap (at around the M2/M3/M5/M62 metallicity), and there is accordingly 
  a tendency to assign a much lower metallicity for an OoII or Oo-Int object 
  in this domain than should really be the case.

Nevertheless, we note that none of the conclusions on the Oo-Int status of For~4
are affected in any way by a change in the cluster metal abundance from $-2.0$ to $-1.5$~dex.
Indeed, 
for both values of metallicity For~4 
%adopting a metallicity of ${\rm [Fe/H]}=-2.0$ from Buonanno et al. (1999) 
%
%%and in agreement with
%%the [Fe/H] values inferred using the relations of Sandage (2006) from {\rm $\langle Pab\rangle$}
%%the mean RRab period 
%%([Fe/H]=$-1.96$) and from 
%the shortest period RRab 
%%$P_{ab,min}$ 
%%([Fe/H]=$-2.25$), and of Sandage (1993) from 
%the mean RRc period 
%%{\rm $\langle Pc\rangle$} ([Fe/H]=$-1.90$),
%the cluster fits right in the middle of
is found to fall in the Oo-Int region in the
$\langle P_{ab}\rangle - {\rm [Fe/H]}$ diagram (Pritzl et al. 2002;
Catelan 2005).
In particular, 
%{\bf Note that 
adopting a metallicity of 
${\rm [Fe/H]} = -1.5$ according to Strader et al. (2003), would place For~4 
right in the middle of the Oo-Int band in the [Fe/H]-HB type 
diagram (see Fig.~8 in Catelan 2005), 
whereas 
%our currently 
adopting ${\rm [Fe/H]}=-2.0$ actually places the cluster slightly 
to the left
%red 
of this band.
% None of the conclusions based on P$_ab,min$
%are affected in any way by a change in the metallicity.}
%%% (see Pritzl et al. 2002, and Catelan 2007).
%This is because, 
In fact, 
and as discussed in a companion paper (Catelan et al. 2007, in preparation), the key 
quantities
defining Oosterhoff status appear to be the average and minimum
   periods of the {\it ab-}type RR Lyrae stars. In terms of both these
   quantities, For~4 appears entirely consistent with an
   Oo-Int classification. This
is also what is generally expected on the basis of the position of
the cluster in the HB type-metallicity plane.
% (see Fig.~8 in Catelan
%2005). 
Note, in this sense, that our photometry and preliminary
number counts suggest a revised Lee-Zinn HB type of
$(B\!-\!R)/(B\!+\!V\!+R) \approx -0.75$ for the cluster---where $B$,
$R$, and $V$ represent the numbers of blue, red, and variable stars
on the HB, respectively.
The Oo-Int behavior further confirms the broad similarity of For~4 to the MW cluster Rup~106.
Indeed, assuming   
$\langle P_{ab}\rangle=0.616$~d (Kaluzny, Krzeminski, \& Mazur 1995) and ${\rm [Fe/H]}=-1.67$ 
(Harris 1996), Rup~106 is one of the few Galactic GC's falling in the 
Oosterhoff-intermediate band in the [Fe/H]-HB type diagram.
% (see Fig.~8 in Catelan 2005). 
However, there are some differences between the two clusters at a more detailed level: for instance,
Rup~106 has no RRc whatsoever, while For~4 has a few; and, if we adopt a boundary of 0.62~d between 
Oo-Int and OoII clusters (Catelan 2004b, 2005), Rup~106 is in fact a ``borderline" Oo-Int cluster,
whereas For~4 falls closer to the middle of the Oo-Int zone.

\section{Cluster distance}
 
The average apparent magnitude of the RR Lyrae stars 
within $30\arcsec$ from the For~4 center (and hence more likely cluster
members), and excluding star 
V14 which may be a background object, is $\langle V(RR)\rangle=21.43 \pm 0.03$~mag  
($\sigma=0.10$~mag, average over 12 stars)
and $\langle B(RR)\rangle=21.76\pm 0.04$~mag ($\sigma=0.13$~mag, average over 12 stars).
%We have excluded also star 
%V14, since it may be a background object. 
Average values become instead $\langle V(RR)\rangle=21.38\pm 0.02$~mag  ($\sigma=0.10$~mag, 
average over 19 stars)
and $\langle B(RR)\rangle=21.74\pm0.03$~mag ($\sigma=0.11$~mag, average over 19 stars)
if we consider ``all" the variable stars in the field of For~4 but V14. 
Both these values, and the latter in particular, are consistent with 
%{\rm $\langle V\rangle$} value is in excellent agreement with
the value of $\langle V(HB)\rangle=21.37 \pm 0.04$~mag, for
the mean level of the (red) old HB stars in the Fornax field, as derived
by Saviane, Held, \& Bertelli (2000).
On the other hand, our values are respectively 0.09 and 0.14 mag brighter 
than $V_{HB}=21.52 \pm 0.05$~mag found by Buonanno et al. (1999)
using HB stars within a distance of $18\arcsec$ from the For~4 center.
The 0.09~mag  difference remains even if we consider 
%the average luminosity of 
%the average luminosity of 
only the 8 RR Lyrae stars with reliable photometry, located 
%in the same area of 
%the cluster covered by Buonanno et al. (1999)
within $18\arcsec$ from the cluster center. 
Unfortunately, we lack reliable photometry for a further 9, more internal,  
RR Lyrae stars which are within $8\arcsec$ from the cluster
center.
% we obtain 
%{\rm $\langle V\rangle(RR)$}=21.43XXXXX$\pm$0.03XXXX mag ($\sigma$=0.10XXXX mag, average on 12XXXX stars)
%
%We lack reliable photometry for the 9 RR Lyrae stars in the
%most central region of For~4 (i.e. for variable stars within $8\arcsec$ from the cluster
%center)
%however can restrict our comparison with  Buonanno et al. (1999)
%to the RR Lyrae stars  compare if only consider compare can compare 
% to make a direct comparison 
%of average luminosities of RR Lyrae and  HB non variable stars
%in the same area of For~4 covered by Buonanno et al. (1999). 
We note that 
a $\sim 0.1$~mag difference between our 
$\langle V(RR)\rangle$ and Buonanno et al.'s $V_{HB}$ might easily be
accounted for if the For~4 RR Lyrae stars were more evolved than their
non-variable HB counterparts, as observed in a number of the metal-poor 
GC's in the MW.  Unfortunately, unlike those GC's, For~4 has a 
predominantly {\em red} HB, although it should be noted that the Catelan (1993) 
synthetic HB's for a predominantly red HB population also predict a difference 
in luminosity, at a fixed temperature, between the lower envelope 
of the RR Lyrae distribution on a CMD and the zero-age HB luminosity level
(see his Table~1, last row, and also Fig.~3 in Catelan 1992). 

In order to derive the distance to For~4 from the average luminosity 
of its RR Lyrae stars we need an estimate of the cluster reddening
and metallicity, as well as values for the slope and zero point 
%absolute magnitude and for the
%slope 
of the  RR Lyrae luminosity-metallicity relation.
%To estimate the distance to For~4, 
We have assumed an absolute magnitude of $M_V = 0.59\pm0.03$ for RR Lyrae stars of 
${\rm [Fe/H]} =-1.5$ (Cacciari \& Clementini 2003; Catelan 2005),
and $\Delta M_V/\Delta{\rm [Fe/H]}=0.214\pm 0.047$ (Clementini et al. 2003, Gratton et 
al. 2004) for the slope of the luminosity-metallicity relation. 
%%For the cluster reddening, $E(\bv)$= Buonanno et al. (1999) reports $E(V-I)$=0.15 $\pm$ 0.06 (transformed
%%to $E(\bv)$ using Cardelli et al. 1989 relation $E(\bv)$=1.31$\times E(V-I)$)

%For the reddening 
We then adopt a standard extinction law [$A_V=3.1\times E(\bv)$] and a reddening value 
of $E(\bv)=0.10 \pm 0.02$ for the cluster, which is the
weighted average of 
%Buonanno et al. (1999) $E(V-I)$=0.15 $\pm$ 0.06 (transformed
%to $E(\bv)$ using Cardelli et al. 1989 relation 
%%$E(V-I)$=1.31$\times E(\bv)$),
$E(\bv)=0.12 \pm 0.05$ from Mackey \& Gilmore (2003a) and $E(\bv)=0.08 \pm 0.03$, the reddening 
we find by matching  
%adopt $E(\bv)$=0.10$\pm 0.04$ (Buonanno et al. 1999), which is also consistent
%%is the color shift required to match
 the blue edge of the RR Lyrae strip
in M3 [$E(\bv)_{\rm M3}=0.01$; Harris 1996] to the colors of the bluest RR Lyrae stars
in For~4.  

%%{\bf Intrinsic $(\bv)_0$ 
%%colours and absolute magnitudes (M$_V$), 
%
%effective temperatures (T$_{eff}$), 
%%were also computed from the Fourier parameters of the light curves
%%of the RRab variables that satisfy the {\it compatibility conditions} 
%%using equation (6) of Kov\'acs \& Walker (2001) with zero points from
%%Kov\'acs \& Jurcsik (1997), and 
%%eq. (1) of Kov\'acs (2002) with the zero point set at 
%%$M_V$=0.59$\pm$0.03 for [Fe/H]=$-1.5$ (Cacciari \& Clementini 2003).
%of the
%Fourier decomposition of the $V$ light curves applying 
%%Jurcsik \& Kov\'acs (1996), Kov\'acs and Walker (2001) and
%JK96, KW01 and 
%%Kov\'acs (2002).
%; hereinafter K02).
%%They are provided in Columns 3 and 5 of Table~6.
%%The $(\bv)_0$ colours derived from the Fourier parameters
%%were compared with the observed $(\bv)$ values (Column 
%%6 of Table~6) to derive
%%constraints on the cluster reddening. We find that ... on average ....thus
%%confirming our assumption of $E(\bv)$=0.10$\pm$0.02 mag (CHECK).}

For ${\rm [Fe/H]}=-2.0$ 
(Buonanno et al. 1999), the cluster's distance modulus is then $20.64 \pm 0.09$~mag, while it 
becomes $20.53 \pm 0.09$~mag if the Strader et al. (2003) metallicity is adopted.
Here, errors are the sum in quadrature of uncertainties of 0.03~mag in $\langle V(RR)\rangle$
(dispersion of the average),
0.05~mag in the zero point of the photometry, 0.02~mag in $E(\bv)$ (corresponding to 0.06~mag in
$A_V$), and of 0.03~mag and 0.047~mag/dex, respectively, in the zero point and in the slope of the 
RR Lyrae luminosity-metallicity relation. 
%%{\bf We list in Column 4 of Table 6 the absolute magnitudes consistent with these distance moduli
%%for the RRab stars satisfying the {\it compatibility conditions}. These $M_V$ values can be 
%%compared
%%to the absolute magnitudes obtained from the Fourier parameters (Column 3 of Table 6).
%%We find that .........}
%%

Previous distance estimates for the Fornax dSph field range from
 $\mu_0=20.59\pm 0.22$~mag (Buonanno et al. 1985) to 20.76~mag (Demers et al.
 1990; Buonanno et al. 1999, for an assumed metal abundance of the Fornax
 field stars of $-1.4$). Saviane et al (2000) infer 
 $\mu_0=20.70\pm 0.12$~mag from the tip of the field stars' red giant branch and
 $\mu_0=20.76\pm 0.04$~mag from the mean magnitude of Fornax's old HB 
 stars, which are assumed to have ${\rm [Fe/H]}\sim -1.8$ and $E(\bv)=0.03$~mag.
 Buonanno et al. (1999) derive an average distance 
 modulus of $\mu_0=20.62\pm 0.08$~mag for the Fornax clusters \#1, 2, 3, and 5 and,
 more recently, Mackey \& Gilmore (2003a) find for the same clusters 
 distance moduli 
 %for the other four globular clusters in Fornax to be
 in the range from  $20.58\pm0.05$ to $20.74\pm0.05$~mag, with an average value of
 $\mu_0=20.66\pm 0.03$ (random) $\pm 0.15$ (systematic) mag.
%Since different assumption about reddening, metallicity and RR Lyrae luminosity-metallicity
% relation are assumed by the different authors, 

For ease of comparison we have summarize in Table~5 the
various distance determinations and the different assumptions about  
reddening, metallicity and RR Lyrae luminosity-metallicity
relation they are based on.
%% 
%%%%({\it Comment del referee:
%%%%
%%%%There is no comparison of their distance value with other values in the 
%%literature for Fornax; there should be.  With the greater metallicity their 
%%distance to For~4 changes to 20.69 mag. The authors note that their mean RR 
%%Lyr V mag agrees extremely well with the mean HB V value from Saviane et al 
%%(2000, A\&A, 355,56); but they don't mention that Saviane et al infer a distance 
%%modulus from the HB of 20.76+/-0.04 {\it (Qui probabilmente il problema
%%e' l'Mv assunta per il braccio orizzontale, vedere cosa assumono Saviane et al.,
%%e quali sono le altre assunzioni che portano al loro modulopi\'u lungo.}
%%which is markedly different from the result 
%%in this paper prior to changing [Fe/H] to -1.5.  Similarly, Mackey and Gilmore 
%%find distance moduli for the other four globular clusters in Fornax to be 20.58 - 20.74. 
%%With the revised metallicity the authors' distance estimate is in much better agreement
%% with other published values.)  ESPANDERE LA PARTE DELLA DISCUSSIONE SULLA DISTANZA.
%% Qui mi sembra che il referee abbia scritto una cretinata!!!! con la metallicita' piu'
%% alta il modulo di distanza viene piu' corto, non piu' lungo, quindi tutta questa discussione
%% viene sbagliata}
%%
%%
Our ``long" distance modulus of $20.64\pm 0.09$~mag, based on ${\rm [Fe/H]}=-2.0$ and
$E(\bv)=0.10$~mag for For~4,
agrees well with both the 
Saviane et al. (2000) and the Mackey \& Gilmore (2003a) estimates, once differences in the 
adopted  reddening, cluster metallicity, and RR Lyrae luminosity-metallicity relation 
are properly accounted for. It is, however, 0.09~mag shorter than one would 
 infer by assuming the Buonanno et al. (1999) $V_{HB}=21.52 \pm 0.05$~mag  
 for  the cluster. On the other hand,  the ``short" distance modulus of $20.53\pm 0.09$~mag, 
 based on 
 ${\rm [Fe/H]}=-1.5$ for For~4, is significantly shorter than any previous distance determination
 for the Fornax system.
 
 \section[]{Summary and Conclusions}

We have presented the first study of the variable star population in For~4, 
the GC located in the central region of the Fornax dSph galaxy. Our sample includes  
29 variable stars, of which 
27 are RR Lyrae stars (22 fundamental-mode, 3 first overtone, and 2 double-mode pulsators).
Twenty-two of these variable stars are located within $30\arcsec$ from the cluster center, 
hence are very likely cluster members.
The average periods of {\it ab-} and {\it c-}type RR Lyrae stars, 
and the minimum period of the {\it ab-}type pulsators,  
point to an Oo-Int status for the cluster,
unlike what is seen for the vast majority of the Galactic GC's.
A similar Oo-Int classification is suggested by 
Mackey \& Gilmore (2003a) also for the other 4 GC's in Fornax.

The Oo-Int type of For~4 further supports the cluster similarity 
with the unusual ``young halo" Galactic cluster Rup 106, and strengthens the
claim for an extragalactic origin of this MW cluster.
Our results on the Fornax field and cluster variable stars also indicate that 
while galaxies like 
the Fornax dSph unlikely 
contributed large fractions of the Galactic halo, they 
 may have provided instead some of the unusual clusters being observed in the MW
 (see e.g. Catelan 2007).
%%
%%an 
%%Since no Oo-Int type is observed among the MW GC's, 
%, an Oo-type that has no counterpart among
%the Milky Way GC's. {\bf A similar Oo-type classification is also inferred for the
%other 4 GC's in Fornax by Mackey \& Gilmore 2003).}
%%the MW halo cannot have been assembled by Fornax dSph-like protogalactic fragments.
%%
%%While .... la maggior parte dei MW GC's sono divisi nei due gruppi di Oosterhoff
%%ect. etc. alcuni pochissimi sono intermedi tra questi Rup 106 e For e' uguale a
%%Rup 106
%%and further supports the similarity 
%%of For~4 with the unusual ``young halo" Galactic cluster Rup 106
%%che si ritiene appartenesse ad una dSph distrutta dalla MW e i cui framemnti sono 
%%stati catturati dalla MW ...... 

The average luminosity of the RR Lyrae stars that are more likely cluster members is 
$\langle V(RR)\rangle=21.43 \pm 0.03$~mag, leading to distance moduli of   
$\mu_0=20.64 \pm 0.09$~mag or $\mu_0=20.53 \pm 0.09$~mag, depending on  
whether Buonanno et al.'s (1999) ${\rm [Fe/H]}=-2.0$ or Strader et al.'s (2003) 
${\rm [Fe/H]}=-1.5$ determination is  
adopted for the cluster metallicity, and for an assumed reddening $E(\bv)=0.10$~mag. 
The properties of the cluster RR Lyrae
stars 
%with more certain membership 
also support a low metal abundance for For~4 closer to 
Buonanno et al.
value, and a distance modulus of $\mu_0=20.64 \pm 0.09$~mag
($D=134\pm 6$~kpc).
This distance modulus agrees well with both Saviane et al. (2000) and
Mackey \& Gilmore (2003a), once differences in the 
adopted  reddening, cluster metallicity, and adopted RR Lyrae luminosity-metallicity 
relation are properly taken into account and corrected for. 

Mackey \& Gilmore (2003a) find that the distance moduli of the Fornax clusters \#1, 2, 3 and 5
are consistent with a line of sight depth of $\sim 8- 10$~kpc for this galaxy.
The distance modulus of For~4 and our preliminary results from the study of the RR 
Lyrae stars in clusters \#2, 3 and 5 show that For~4 is at the center of the Fornax GC's 
distance moduli distribution, and 
confirm  Mackey \& Gilmore's finding of a line-of-sight depth in Fornax, although  
reducing its extent to about 7 kpc. Further insight on the line of sight depth of the
Fornax dSph will be gained from our study of the galaxy's field variable stars.

\acknowledgments 
We thank the anonymous referee for comments and suggestions that have helped to
improve the paper. This research was funded 
by MIUR, under the scientific project:  2004020323 (P.I.: M. Capaccioli) 
and by 
PRIN INAF 39/2005  
%''Star formation histories of resolved galaxies: the local route to cosmology"
(P.I.: M. Tosi). 
CG  acknowledges  a Marco Polo Fellowship by the University of Bologna.
MC acknowledges support by Proyecto FONDECYT Regular No. 1071002.
HAS thanks the National Science Foundation for support under grant AST 0607249.
%-----------------------------------------------------------------
%-----------------------------------------------------------------
%-----------------------------------------------------------------
%-----------------------------------------------------------------

\clearpage

\appendix
\section{Atlas of the Light Curves}

%{Atlas of the light curves}                          

Atlas of light curves for the 29 variable stars identified in For~4.
Photometric data 
are folded according to the ephemerides provided in Table~3.
Variable stars are ordered by increasing distance from the cluster center.
Only differential flux light curves from the Magellan $V$ dataset are 
available for stars from V1 to V9. 

%\clearpage

\begin{table*}
   \fontsize{6}{6}
      \caption[]{Instrumental set-ups and logs of the observations}
	 \label{t:obs}
     %%%%%%%%%%%%%%%$$
	 %%%%%%%%%%%%%%%%%%%%\begin{array}{llcc}
	 $$
	\begin{array}{llllllrrc}
	   \hline
	    \hline
	   \noalign{\smallskip}
{\rm ~~~~~Dates}		   & {\rm ~~~~Telescope}     & {\rm ~~Instrument} & {\rm ~~~Detector}			 
&{\rm ~Resolution}  & {\rm ~~~~FOV}	&{\rm N_ B} &{\rm N_ V}& {\rm Photometric~precision}\\
	         &			  &		     &      &     &  & & & {\rm (HB~level)}\\
{\rm ~~~~~~UT}   &			  &		     &~~~~{\rm (pixel})&{\rm ~~(\prime \prime/pixel)}&  & & & {\rm (mag)}\\
	    \noalign{\smallskip}
	    \hline
	    \noalign{\smallskip}
{\rm Nov. 13-15, 2003} & {\rm ~~Magellan/Clay} & {\rm  ~~~MagIC    }  &{\rm 2048\times2048~SITe}&~~~0.069&2.35^{\prime}\times\ 2.35^{\prime}& 10~&39~& 0.01\\
{\rm Dec. 1-2, 2004}   & {\rm ~~Magellan/Clay} & {\rm  ~~~MagIC    }  &{\rm 2048\times2048~SITe}&~~~0.069&2.35^{\prime}\times\ 2.35^{\prime}& 9~&19~& 0.01\\
{\rm Nov. 14, 2003}    & {\rm ~~Blanco/CTIO}	& {\rm  ~~~Mosaic II}  &{\rm 2048\times4096~SITe}&~~~0.27&~36^{\prime}\times\ 36^{\prime} &  8~&16~& 0.03-0.05\\
{\rm Oct. 25-26, 2004} & {\rm ~~Blanco/CTIO}	& {\rm  ~~~Mosaic II}  &{\rm 2048\times4096~SITe}&~~~0.27&~36^{\prime}\times\ 36^{\prime} & 51~ &124~& 0.03-0.05 \\
{\rm Sept. 9, 2005}    & {\rm ~~Blanco/CTIO}	& {\rm  ~~~Mosaic II}  &{\rm 2048\times4096~SITe}&~~~0.27&~36^{\prime}\times\ 36^{\prime} & 2~& 2~& 0.03-0.05\\
\hline
	  %%%%%%%%%%%%%%  \end{array}
	 %%%%%%%%%%%   $$
	 \end{array}
	 $$
	    \end{table*}

\begin{table}
\begin{center}
\caption{$V,B$ photometry of the For~4 variable stars with light curves in magnitude scale}
\vspace{0.5 cm} 
%\small
\begin{tabular}{cccc}
\hline
\hline
%\multicolumn{1}{c}{}& \multicolumn{1}{c}{}&
\multicolumn{4}{c}{Star V18 - {\rm RRab}} \\
%\begin{tabular}{cccc}
%\hline
%\noalign{\smallskip}
%& & Star V18 - {\rm RRab}&  \\
\hline
{\rm HJD} & {\rm V}  & {\rm HJD } & {\rm B}\\ 
{\rm ($-$2452956)} &   & {\rm ($-$2452956) } &  \\
%            \noalign{\smallskip}
\hline
%\noalign{\smallskip}
     0.697471 &    21.86  &       0.749611 &    22.26 \\
     0.702402 &    21.79  &       1.563541 &    21.66 \\
     0.708871 &    21.76  &       1.604275 &    21.93 \\
     0.714947 &    21.79  &       1.631649 &    21.97 \\
     0.729866 &    21.77  &       1.665676 &    22.10 \\
     0.735942 &    21.83  &       1.690539 &    22.18 \\
     0.742018 &    21.83  &       1.717479 &    22.24 \\
     0.757122 &    21.84  &       1.744663 &    22.33 \\
     0.763198 &    21.84  &       1.772030 &    22.31 \\
     0.784262 &    21.84  &       2.554781 &    20.88 \\
\hline

\end{tabular}

%\label{t:table1}
\end{center}
\medskip

A portion of Table 2 is shown here for guidance regarding its form
and content. The entire catalog is available in the electronic edition 
of the journal.
\end{table}
\normalsize

\begin{table}
\tiny
\caption[]{Identification and properties of the For~4\ \ variable stars}
\label{t:fornax4}
     $$
         \begin{array}{lrcclllclclccc}
	    \hline
            \hline
           \noalign{\smallskip}
           {\rm Name} & {\rm Id}  & {\rm \alpha } & {\rm \delta} &  {\rm Type} &~~~{\rm P} & 
	    ~~~{\rm Epoch}  & {\rm \langle V\rangle}  & ~~{\rm N_V} & {\rm \langle B\rangle}  &~~{\rm N_B} &
	    {\rm A_V} & {\rm A_B} & {\rm Notes}\\
            ~~{\rm (a)}& &{\rm (2000)}& {\rm (2000)}& & ~{\rm (days)}& ($-$2450000) & & & 
	    & &  & & \\
            \noalign{\smallskip}
            \hline
            \noalign{\smallskip}
	    
{\rm ~V1}&{\rm ~LC}378 & 2:40:07.62 & -34:32:09.4 & {\rm RRab}  & 0.54::  &  $\nodata$    &$\nodata$ &   54  &$\nodata$ & $\nodata$   & $\nodata$  & $\nodata$ & {\rm (b,c)}\\
{\rm ~V2}&{\rm ~LC}360 & 2:40:07.70 & -34:32:09.8 & {\rm RRab}  & 0.6261  &  2955.77      &$\nodata$ &   52  & $\nodata$ & $\nodata$   & $\nodata$  & $\nodata$ & {\rm (b)}\\
{\rm ~V3}&{\rm ~LC}330 & 2:40:07.47 & -34:32:10.7 & {\rm RRab}  & 0.5968  &  3342.640     &$\nodata$ &   56  & $\nodata$ & $\nodata$   & $\nodata$  & $\nodata$ & {\rm (b,c)}\\
{\rm ~V4}&{\rm ~LC}386 & 2:40:07.75 & -34:32:09.0 & {\rm RRab}  & 0.65::  &  $\nodata$    &$\nodata$ &   53  & $\nodata$ & $\nodata$   & $\nodata$  & $\nodata$ & {\rm (b,c)}\\
{\rm ~V5}&{\rm ~LC}319 & 2:40:07.80 & -34:32:11.0 & {\rm RRd?}  & 0.46::  &  $\nodata$    &$\nodata$ &   57  & $\nodata$ & $\nodata$   & $\nodata$  & $\nodata$ &{\rm (b)}\\
{\rm ~V6}&{\rm ~LC}381 & 2:40:07.33 & -34:32:09.2 & {\rm RRab}  & 0.52936 &  2956.736     &$\nodata$ &   55  & $\nodata$ & $\nodata$   & $\nodata$  & $\nodata$ & {\rm (b)}\\
{\rm ~V7}&{\rm ~LC}178 & 2:40:07.49 & -34:32:15.1 & {\rm RRab}  & 0.5::   &  $\nodata$    &$\nodata$ &   55  & $\nodata$ & $\nodata$   & $\nodata$  & $\nodata$ &{\rm (b,c)}\\
{\rm ~V8}&{\rm ~LC}486 & 2:40:07.40 & -34:32:05.1 & {\rm RR?}   & 0.83::  &  $\nodata$    &$\nodata$ &   57  & $\nodata$ & $\nodata$   & $\nodata$  & $\nodata$ & {\rm (b,c)}\\
{\rm ~V9}&{\rm ~LC}291 & 2:40:07.99 & -34:32:11.5 & {\rm RR?}   & 0.69::  &  $\nodata$    &$\nodata$ &   57  & $\nodata$ & $\nodata$   & $\nodata$  & $\nodata$ & {\rm (b,c)}\\
  	 &	       &	    &		  &	        &	  &	        &        &       &         &    &      &   &            \\
{\rm ~V10 } & 2658 & 2:40:07.34 & -34:32:03.6     & {\rm RRab}  & 0.64450 & ~~2956.740  & 21.31 & 51     & 21.75 & 19   & 0.40 & 0.49 & \\
{\rm ~V11}  & 2051 & 2:40:08.39 & -34:32:09.7     & {\rm RRab}  & 0.53085 & ~~3342.568  & 21.42 & 57+136 & 21.80 & 17+60& 1.06 & 1.36 & \\
{\rm ~V12}  & 2272 & 2:40:07.99 & -34:32:20.5     & {\rm RRab}  & 0.59506 & ~~3342.545  & 21.35 & 58+139 & 21.74 & 19+61& 0.83 & 1.03 & \\
{\rm ~V13 } & 2949 & 2:40:06.81 & -34:32:16.9     & {\rm RRab}  & 0.58767 & ~~2958.620  & 21.57 & 52     & 21.79 & 13   & 0.85 & ~~0.75:: & {\rm (d)} \\
{\rm ~V14}  & 1956 & 2:40:08.55 & -34:32:12.4     & {\rm RRab}  & 0.64520 & ~~3303.680  & 22.04 & 55+109 & 22.27 & 19+47& 1.14 & 1.40 &  \\
{\rm ~V15}  & 2076 & 2:40:08.33 & -34:32:19.8     & {\rm RRab}  & 0.67000 & ~~2958.640  & 21.45 & 58+139 & 21.96 & 18+58& 0.27 & 0.42 & \\
{\rm ~V16}  & 2651 & 2:40:07.35 & -34:31:55.7     & {\rm RRab}  & 0.51910 & ~~3341.555  & 21.49 & 57     & 21.80 & 17   & 0.64 & 0.83 & \\
{\rm ~V17 } & 2664 & 2:40:07.33 & -34:32:25.9     & {\rm RRc}   & 0.31618 & ~~2958.573  & 21.30 & 57+141 & 21.55 & 17+63   & 0.53 & 0.67 & \\
{\rm ~V18 } & 3177 & 2:40:06.38 & -34:32:15.7     & {\rm RRab}  & 0.56297 & ~~2958.5475 & 21.44 & 58+124 & 21.79 & 19+39& 1.15 & 1.54 & \\
{\rm ~V19 } & 3303 & 2:40:06.14 & -34:32:08.8     & {\rm RRab}  & 0.52651 & ~~2958.600  & 21.55 & 58+136 & 21.90 & 19+52& 1.20 & 1.50 & \\
{\rm ~V20} & 1685 & 2:40:09.09 & -34:32:17.2      & {\rm RRc}   & 0.36479 & ~~3342.544  & 21.54 & 52+141 & 21.87 & 18+51& 0.40 & 0.43 & \\
{\rm ~V21} & 2948 & 2:40:06.80 & -34:32:33.9      & {\rm RRc}   & 0.39962 & ~~3341.663  & 21.30 & 56+131 & 21.62 & 19+50& 0.36 & 0.42 & \\
{\rm ~V22} & 3544 & 2:40:05.68 & -34:32:20.4      & {\rm RRab}  & 0.65255 & ~~3341.540  & 21.42 & 57     & 21.83 & 19+49& 0.57 & 0.71 & \\
{\rm ~V23} & 1130 & 2:40:10.29 & -34:32:12.4      & {\rm RRab}  & 0.55746 & ~~3342.600  & 21.28 & 56+136 & 21.76 & 19+56& 0.54 & 0.69 & \\
{\rm ~V24} & 4153 & 2:40:04.39 & -34:32:40.9      & {\rm RRab}  & 0.61631 & ~~2958.695  & 21.30 & 58+140 & 21.66 & 19+59& 0.69 & 0.90 & \\
{\rm ~V25} & 2219 & 2:40:07.91 & -34:33:07.2      & {\rm RRd}   & 0.41522 & ~~2958.570  & 21.31 & 56+137 & 21.64 & 19+61& 0.31 & 0.56 & {\rm (e)}\\
% 	   &	  &	       &                  &             & 0.56284 & ~~0.7377    & 	&        &	 &      &      &      & \\	
{\rm ~V26} & 4779 & 2:40:03.13 & -34:31:43.4      & {\rm RRab}  & 0.61610 & ~~2958.600  & 21.31 & 57     & 21.66 & 19   & 1.18   & 1.46 & \\
{\rm ~V27} & 4989 & 2:40:02.63 & -34:32:42.0      & {\rm RRab}  & 0.59643 & ~~3341.620  & 21.36 & 58+140 & 21.60 & 19+61& 0.84 & 1.12 & \\
{\rm ~V28} &  437 & 2:40:11.81 & -34:31:13.6      & {\rm RRab}  & 0.55954 & ~~3303.880  & 21.26 & 55+119 & 21.62 & 17+51& 0.97 & 1.24 & \\
{\rm ~V29} & 5361 & 2:40:01.81 & -34:32:54.7      & {\rm RRab}  & 0.69860 & ~~2958.587  & 21.23 & 58+134 & 21.64 & 19+61& 0.40 & 0.65 & \\
\hline
            \end{array}
	    $$
%\begin{list}{}{}
{\small $^{\mathrm{a}}$ Variable stars were assigned increasing numbers starting from the cluster 
center that was set at $\alpha$~=~02:40:07.6,  $\delta$~=~$-$34:32:10.0 (J2000). Stars from V1 to V22 are
located within $30\arcsec$ from the For~4 center, hence are more likely cluster members.}\\
{\small $^{\mathrm{b}}$ Light curves available in differential flux only.}\\
{\small $^{\mathrm{c}}$ Stars with gaps in the folded light curve.}\\
%%%%%%%{\small $^{\mathrm{d}}$ Candidate double mode RR Lyrae star. The classification was done on the appearance of the
%%%%%%%%%%%light curve that shows two well separate branches. The star power spectrum has two peaks  at
%%%%%%%%%$P_1$= XXX and $P_0$= XXX, respectively ({\bf CHECK with Ennio's e-mail}).}\\
{\small $^{\mathrm{d}}$ Star with the typical period of a fundamental-mode pulsator, but with the light curve of 
a first overtone pulsator.}\\
{\small $^{\mathrm{e}}$ Double-mode \rrl\ star with fundamental-mode period $P_0= 0.56284$~d and period ratio 
$P_1/P_0=0.7377$; in the table we list the star's first overtone period.}\\

\end{table}

\begin{table}
%\tiny
\caption[]{Average quantities for the For~4 RR Lyrae stars}
%\label{t:fornax4}
     $$
         \begin{array}{lrcclllc}
	    \hline
            \hline
           \noalign{\smallskip}
           {\rm N_{RRab}} & {\rm N_{RRc}}  & {\rm N_{RRd}} & {\rm N_{RRc}/N(RR_{tot})} & {\rm \langle P_{ab}\rangle~(d)} &  
	   {\rm \langle P_{c}\rangle~(d)}& {\rm P_{ab,min}~(d)} & {\rm P_{ab,max}~(d)}\\
                    ~~(a)      &      (a)~~          &    (a)           &        (a,b)                &       ~(c)                  &             
                    (d)                 &                  &                  \\
            \noalign{\smallskip}
            \hline
            \noalign{\smallskip}
	    
22(16)& 3(3) & 2(1) & 0.12(0.16)& 0.594 & 0.360  & 0.5191 & 0.6986\\
  &   &   & 0.19(0.20)& 0.596 & 0.374  &        &       \\
\hline
            \end{array}
	    $$
%\begin{list}{}{}
{\small $^{\mathrm{a}}$ Values in parentheses consider only
RR Lyrae stars within $30\arcsec$ from the cluster center.}\\
{\small $^{\mathrm{b}}$ Ratio of RRc to total number of RR Lyrae stars with (2nd raw values)
and without (1st raw values) the RRd stars.}\\
{\small $^{\mathrm{c}}$ Fundamental-mode average period with (2nd raw value) and without
(1st raw value) stars V13 and V14.}\\
{\small $^{\mathrm{d}}$ First overtone average period with (2nd raw value) and without
(1st raw value) the double mode star V25.}\\
\end{table}

\begin{table*}[ht]
%\vskip 1 cm
\begin{center}
\caption{Distance determinations based 
on the HB luminosity of the Fornax dSph field and globular clusters 
}
%\vspace{0.5cm}
\scriptsize
\begin{tabular}{clccccc}
\hline
\hline
%$V_{HB}$&$E(\bv)$&$A_V$ law&[Fe/H]&M$_V${\it vs}[Fe/H]&$\mu_0$&Reference\\ 
$V_{HB}$&$E(\bv)$&[Fe/H]&$M_V$ {\it vs.} [Fe/H]&$\mu_0$&Reference\\ 
 (mag) & (mag)  &       &                   &        & \\
\hline
               (field)& 0.03&                      &              &          20.59$\pm$0.22& Buonanno et al.(1985)\\
21.37$\pm$0.04 (field)& 0.03&         $-1.8$       & 0.17$\times$([Fe/H]+1.5)+0.57&20.76$\pm$0.04& Saviane et al. (2000)\\
               (field)& 0.05&                      &              &          20.76               & Demers et al.(1990)\\
               (field)& 0.05&                      & 0.17$\times$([Fe/H]+1.5)+0.57&20.76$\pm$0.10& Buonanno et al.(1999)\\
21.28$\pm$0.01 (field)&0.04$\pm$0.03&$-1.77\pm0.20$& 0.22$\times$([Fe/H]+1.5)+0.50&20.72$\pm$0.10& Greco et al. (2007)\\
21.20$\pm$0.05 (For 1)&0.04$\pm$0.05&$-2.20\pm0.20$& 0.17$\times$([Fe/H]+1.5)+0.57&20.62$\pm$0.08& Buonanno et al.(1999)\\
21.35$\pm$0.05 (For 2)&0.07$\pm$0.05&$-1.78\pm0.20$&                              &              & Buonanno et al.(1999)\\
21.20$\pm$0.05 (For 3)&0.04$\pm$0.05&$-1.96\pm0.20$&                              &              & Buonanno et al.(1999)\\
21.20$\pm$0.05 (For 5)&0.06$\pm$0.05&$-2.20\pm0.20$&                              &              & Buonanno et al.(1999)\\
21.27$\pm$0.01 (For 1)&0.07$\pm$0.01&$-2.05\pm0.10$& 0.23$\times$([Fe/H]+1.6)+0.56&20.58$\pm$0.05& Mackey \& Gilmore (2003a)\\
21.34$\pm$0.01 (For 2)&0.05$\pm$0.01&$-1.83\pm0.07$&          &20.67$\pm$0.05& Mackey \& Gilmore (2003a)\\
21.24$\pm$0.01 (For 3)&0.04$\pm$0.01&$-2.04\pm0.07$&          &20.66$\pm$0.05& Mackey \& Gilmore (2003a)\\
21.33$\pm$0.01 (For 5)&0.03$\pm$0.01&$-1.90\pm0.06$&          &20.74$\pm$0.05& Mackey \& Gilmore (2003a)\\
21.21$\pm$0.02 (For 3)&0.04$\pm$0.03&$-1.91\pm0.20$& 0.22$\times$([Fe/H]+1.5)+0.50&20.68$\pm$0.11& Greco et al. (2007)\\
                      &             &              &          &              &                         \\
21.43$\pm$0.03 (For~4)&0.10$\pm$0.02&$-2.01\pm0.20$& 0.214$\times$([Fe/H]+1.5)+0.59&20.64$\pm$0.09&This paper\\
\hline
\end{tabular}
\end{center}
%\normalsize
\normalsize
%\label{t:tab7}
\end{table*}

\clearpage

\begin{figure}
\plotone{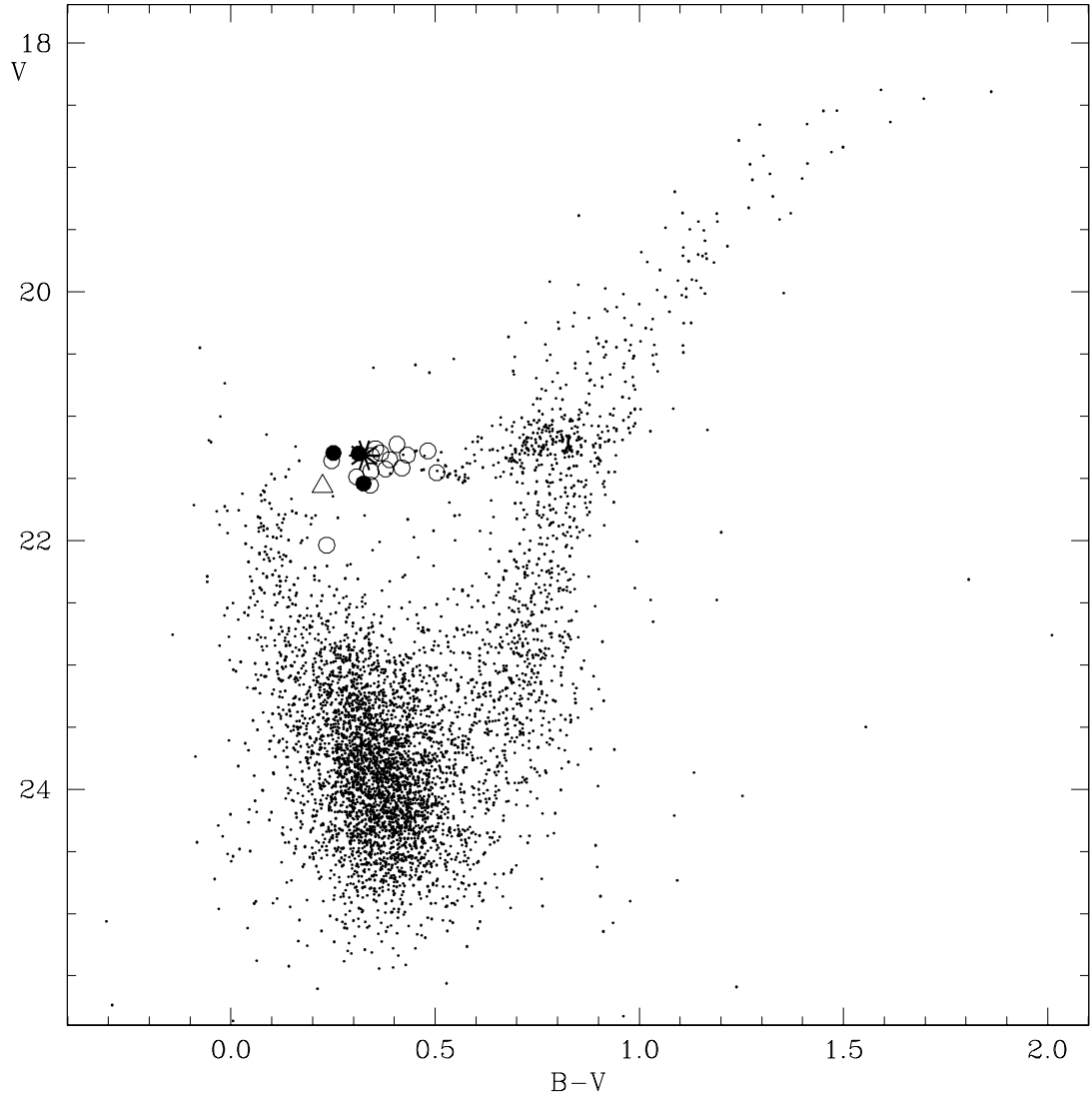}
\caption{Color-magnitude diagram of For~4\ from the Magellan data. 
Variable stars are plotted according to their intensity-averaged magnitudes and colors, using
different symbols for the different types.
{\em Open circles}: {\it ab-}type RR Lyrae stars; {\em filled circles}: first overtone RR Lyrae stars;
{\em asterisk}: double-mode pulsator (RRd). The {\em open triangle} marks a peculiar 
RR Lyrae star (V13).}
%{f:fig2}
\end{figure}

\clearpage
 
\begin{figure}
\plotone{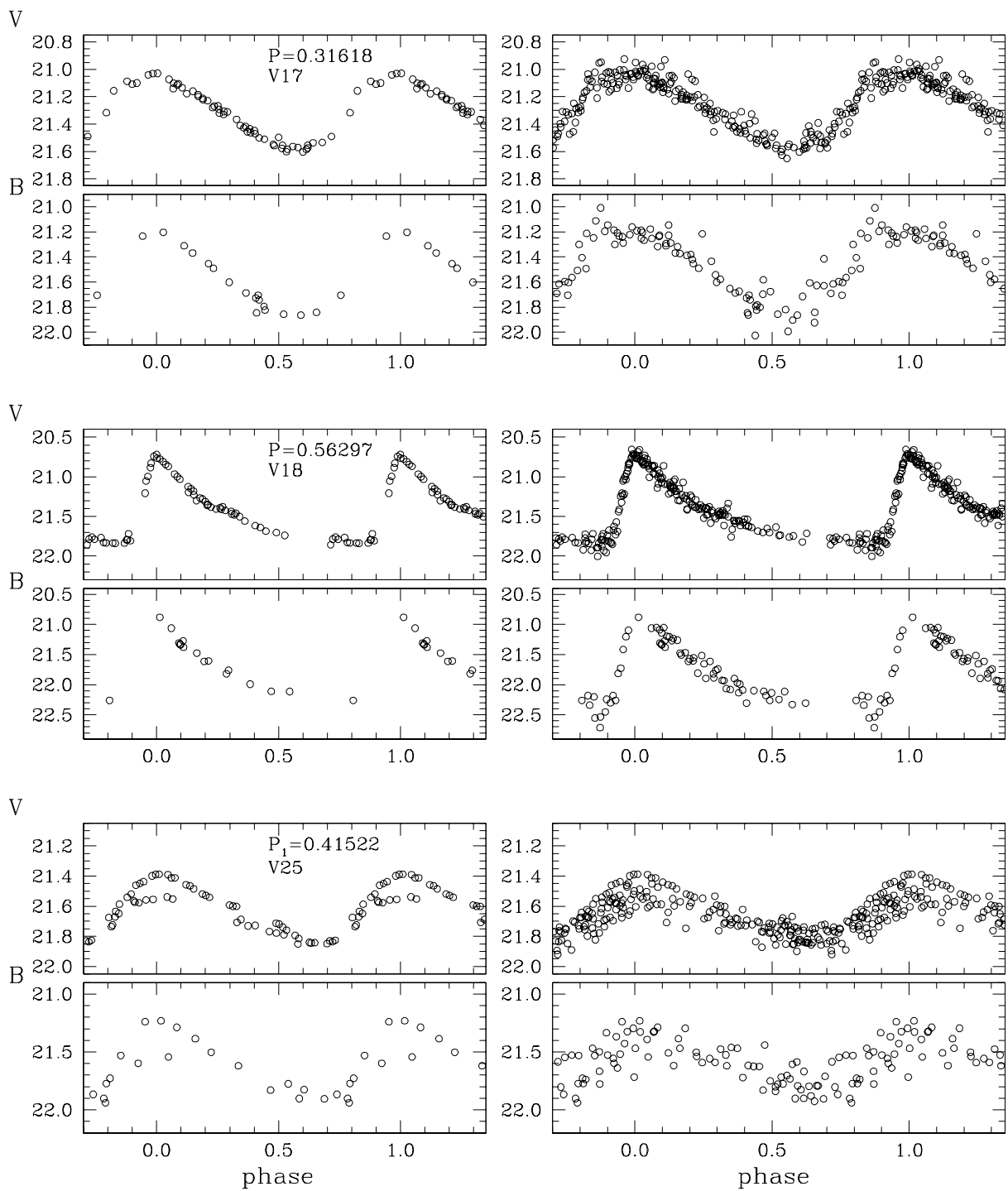}
%\plotone{f2_new.ps}
\caption{$V,B$ light curves of RR Lyrae stars in For~4.
Left panels, from top to bottom: an RRc, an RRab and an RRd star from the Magellan dataset only.
Right panels: the same, for the combined Magellan + CTIO dataset.}
%{f:fig1}
\end{figure}

\clearpage

\begin{figure*}
%%\plottwo{isto.fgc4.ps}{isto.fond.united.ps}
\plottwo{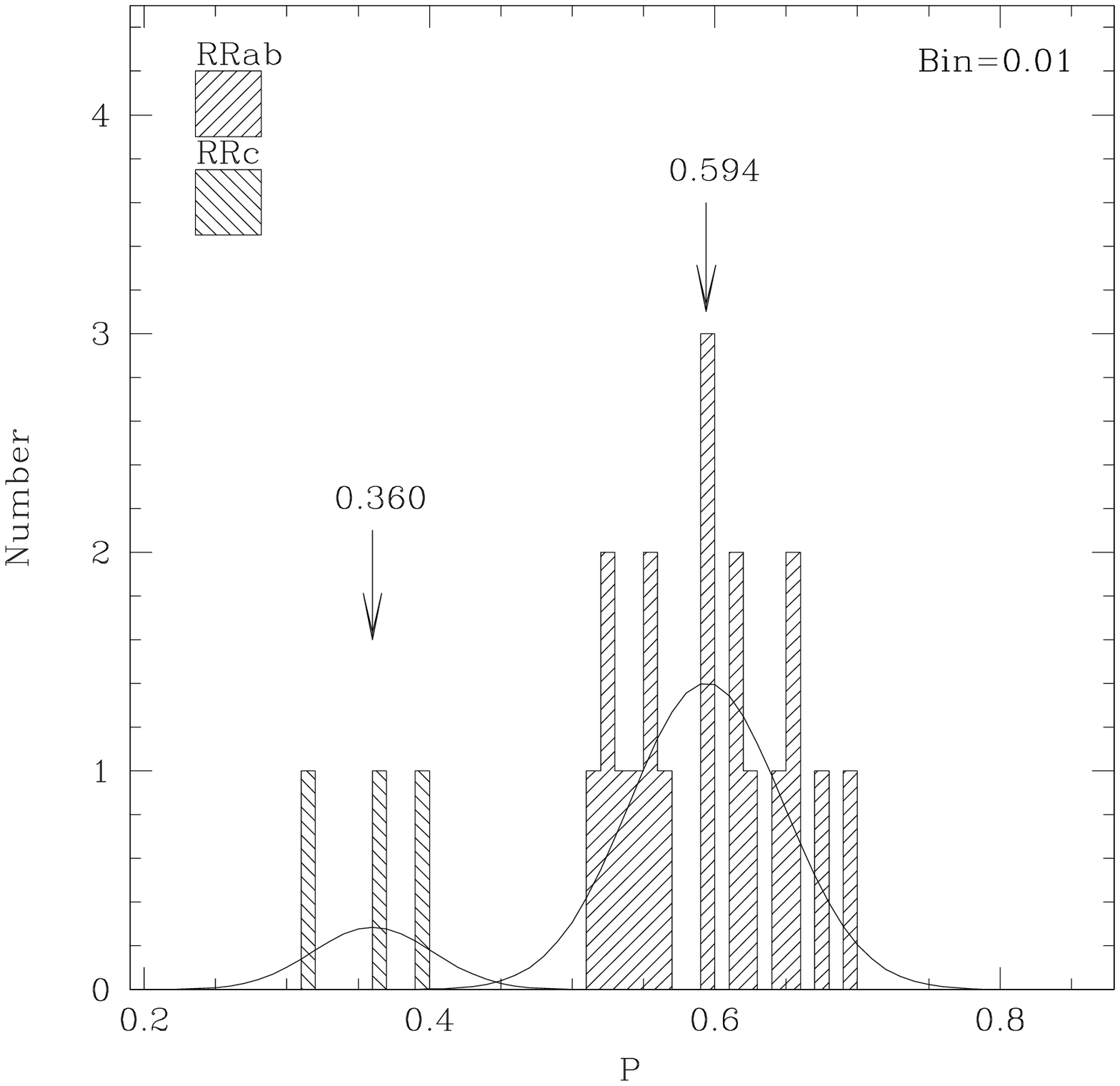}{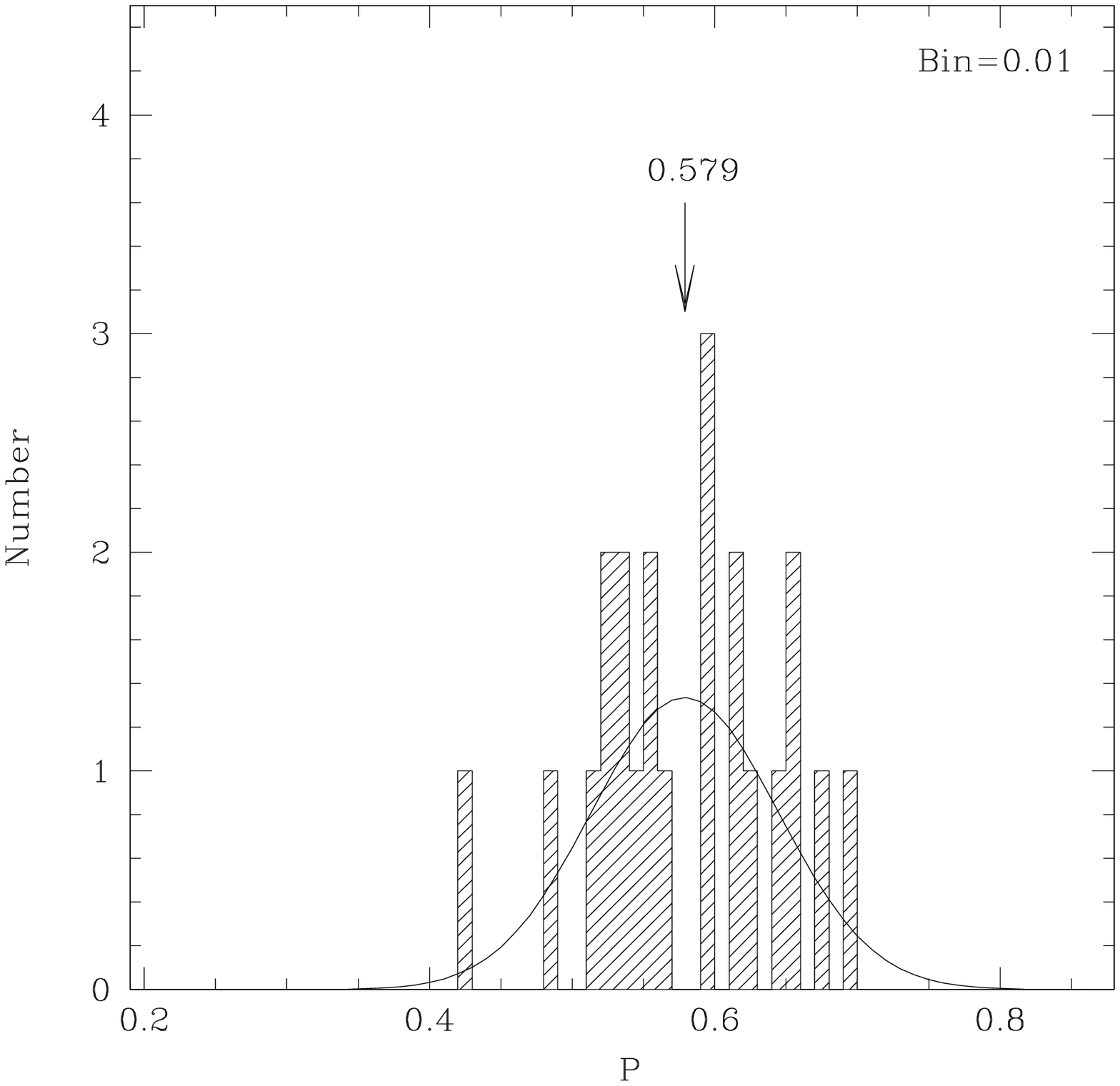}
\caption{{\em Left panel}: Period distribution of fundamental-mode and first overtone \rrl\ stars in For~4.
{\em Right panel}: Fundamentalized period distribution for RR Lyrae stars in For~4. Interestingly, the 
peak of the distribution, at $\approx 0.6$~d, is seemingly located at a longer period than for either 
OoI or OoII GC's. 
%[NOTE: If we decide to keep both panels, the ``mean = 0.579...'' entry in the figure
%should be replaced with ``peak = 0.xyz'' --MC]} 
}
\end{figure*}

\clearpage

\begin{figure}
\plotone{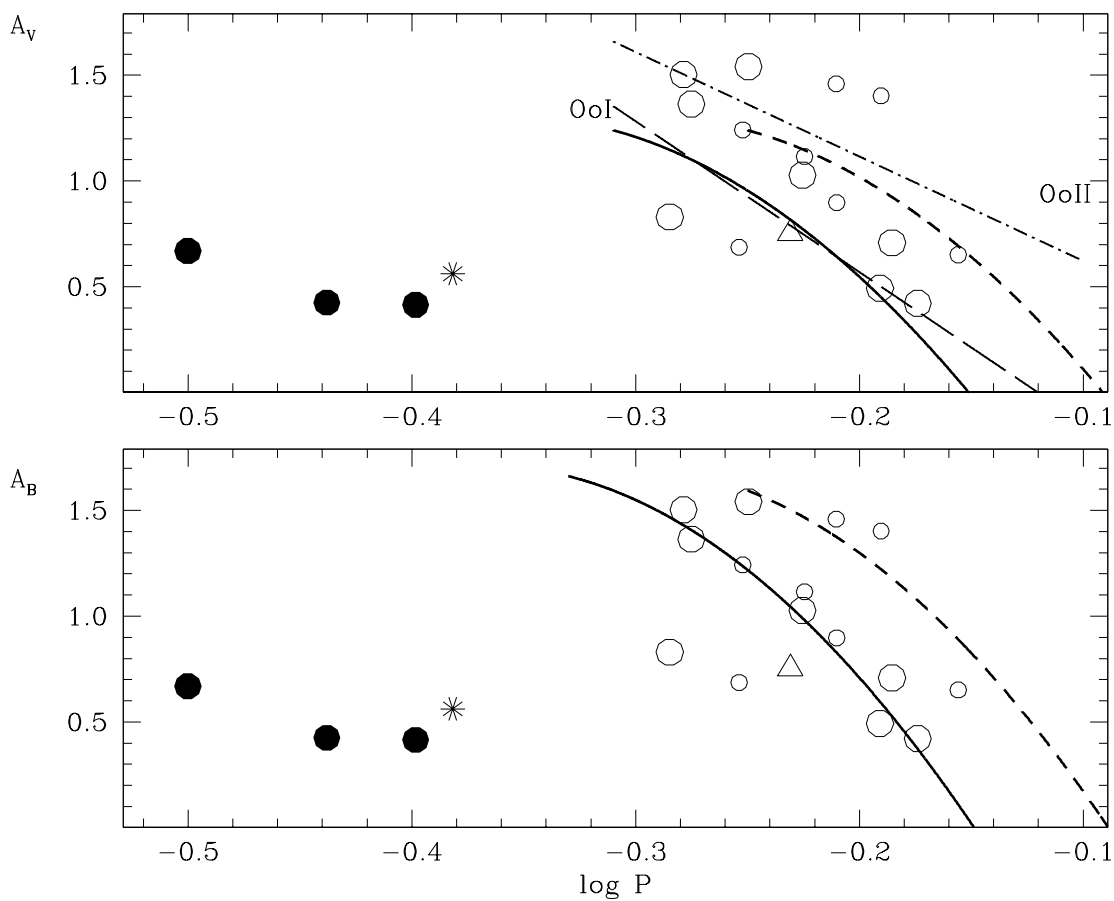}
\caption{$V$, $B$ period-amplitude diagrams of For~4 \rrl\ stars.
% with light curves in magnitude scale.  
Symbols are as in Figure~1, but those indicating variable stars located at distances
less than $30\arcsec$ from the cluster center are expanded.
%Filled and open circles are {\emph c-}
%and {\emph ab-type} \rrl\ stars, respectively. The x sign is V13.
The {\em straight lines}  
%dashed 
are the positions of the OoI and OoII Galactic GC's according to Clement \& Rowe (2000). 
Period-amplitude distributions of the
{\it bona fide} regular ({\em solid curves}) and well-evolved ({\em dashed curves}) {\it ab} \rrl\ stars in M3 
from Cacciari, Corwin, \& Carney (2005) are also shown for comparison.}
%{f:fig3}
\end{figure}

\begin{figure*} 
\includegraphics[width=18cm,bb=40 159 580 703,clip]{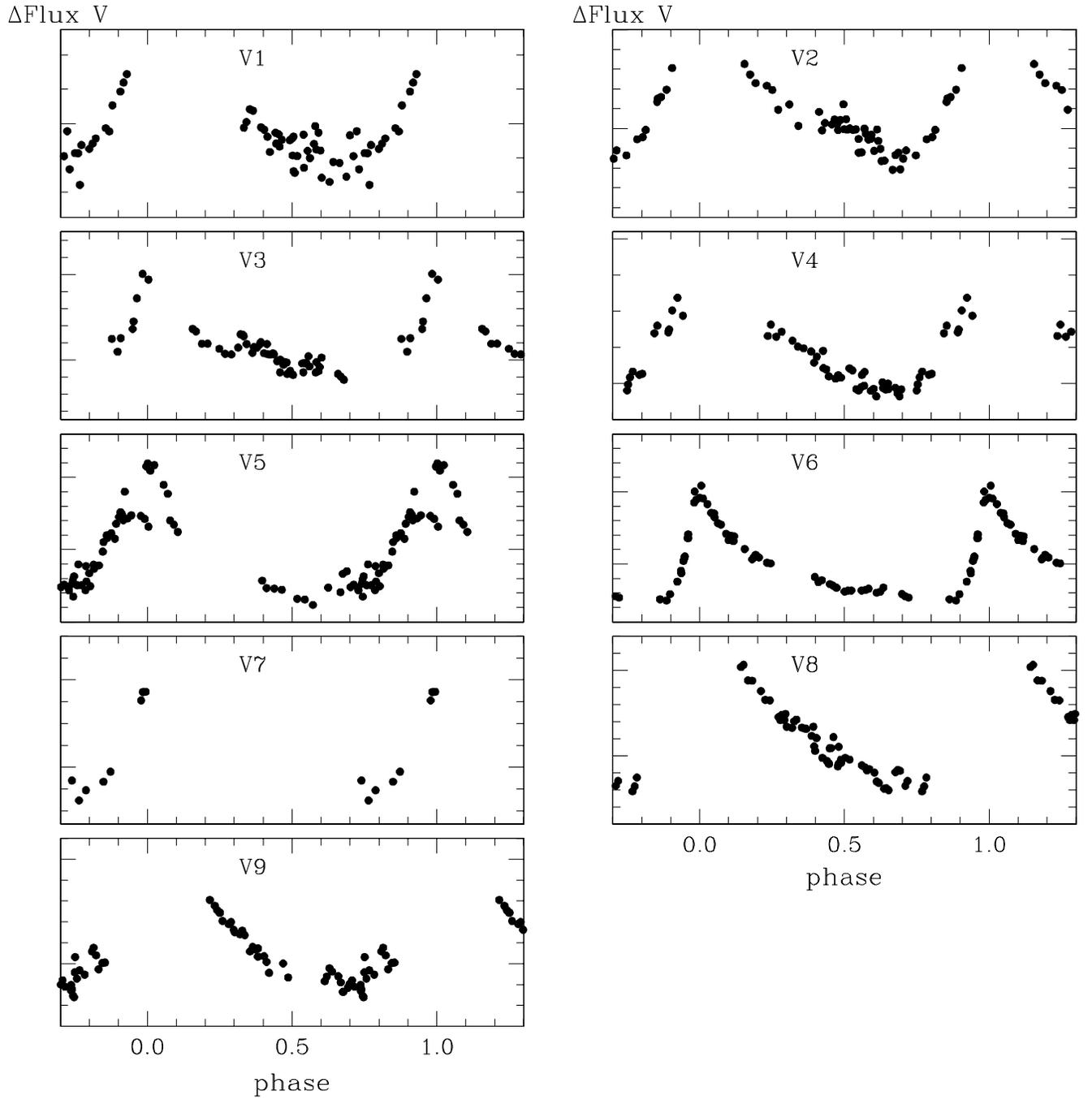}
%%\includegraphics[bb=18 144 592 718, width=18cm, draft=true]{c5.ps}
%\caption[]{Light curves in differential flux, from the Magellan $V$ dataset.}
\caption{Light curves in differential flux, from the Magellan $V$ dataset.}
%\label{f:fig1a}
\end{figure*}

\begin{figure*} 
\includegraphics[width=18cm,bb=40 159 580 703,clip]{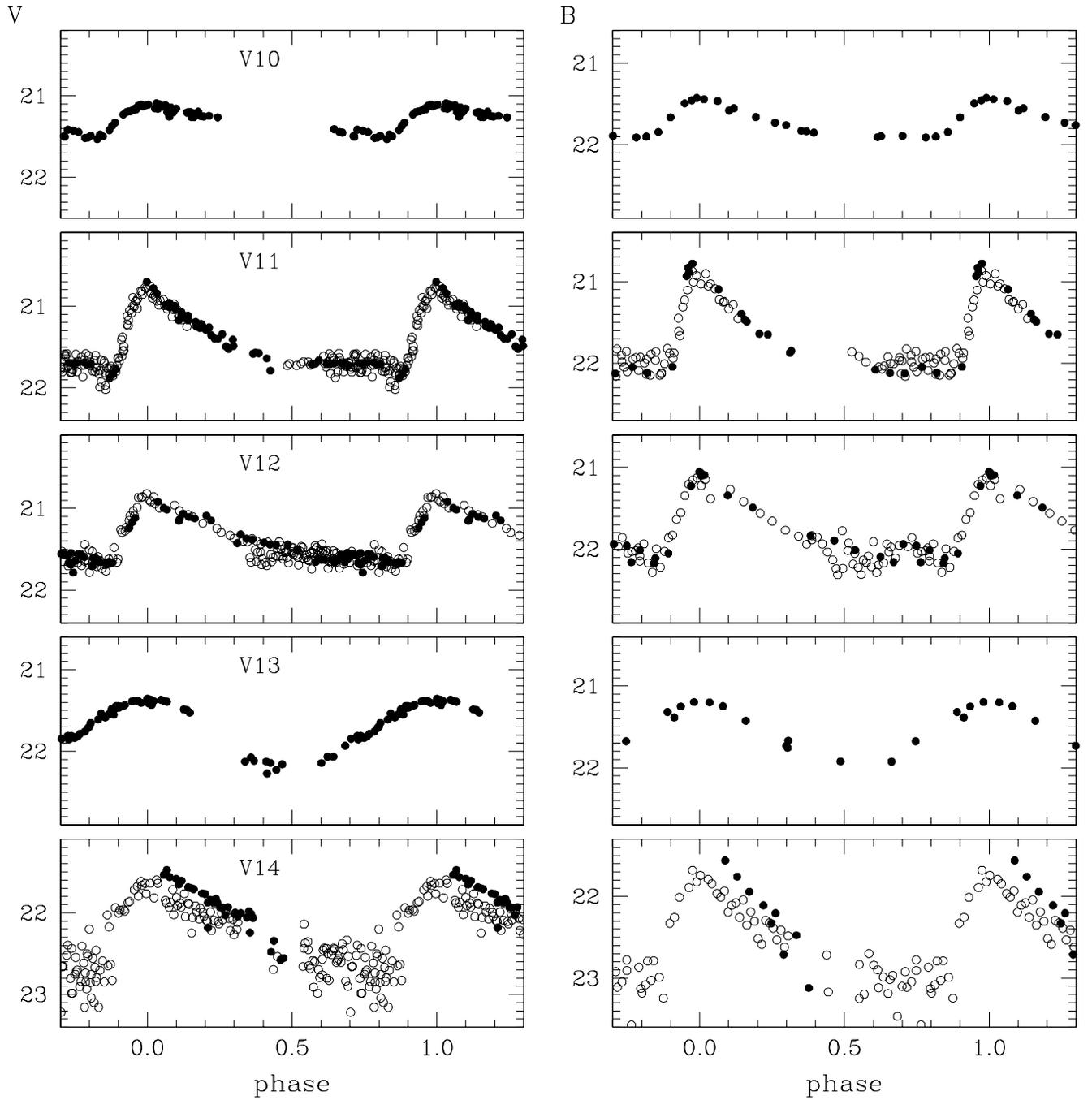}
%\includegraphics[bb=18 144 592 718, width=18cm, draft=true]{difabrizio.figA1_2.jpg}
%%\caption[]{$V$ ({\em left panels}), $B$ ({\em right panels}) light curves
%%from the combined Magellan plus CTIO datasets.}
\caption{$V$ (left panels), $B$ (right panels) light curves
from the combined Magellan ({\em solid circles})  plus CTIO ({\em open circles}) datasets.}
%{Fig. {\bf A.1.} -- continued --}
%\label{f:fig1b}
\end{figure*}

\begin{figure*} 
\figurenum{6}
\includegraphics[width=18cm,bb=40 159 580 703,clip]{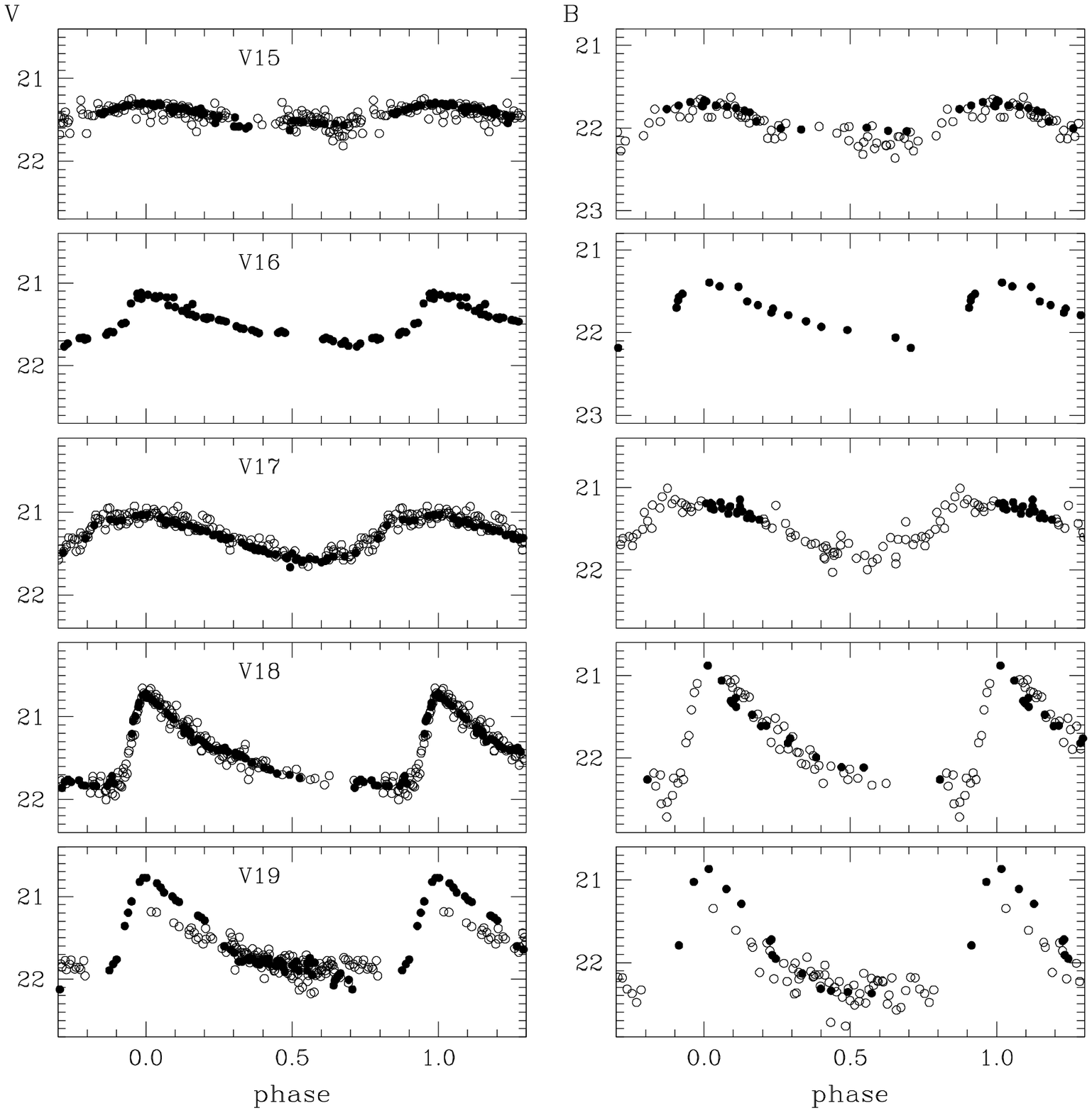}
%\includegraphics[bb=18 144 592 718, width=18cm, draft=true]{difabrizio.figA1_2.jpg}
%%\caption[]{$V$ (left panels), $B$ (right panels) light curves
%%from the combined Magellan (filled circles) plus CTIO (open circles) datasets.}
\caption{continued}
%\label{f:fig1b}
\end{figure*}

\begin{figure*} 
\figurenum{6}
\includegraphics[width=18cm,bb=40 159 580 703,clip]{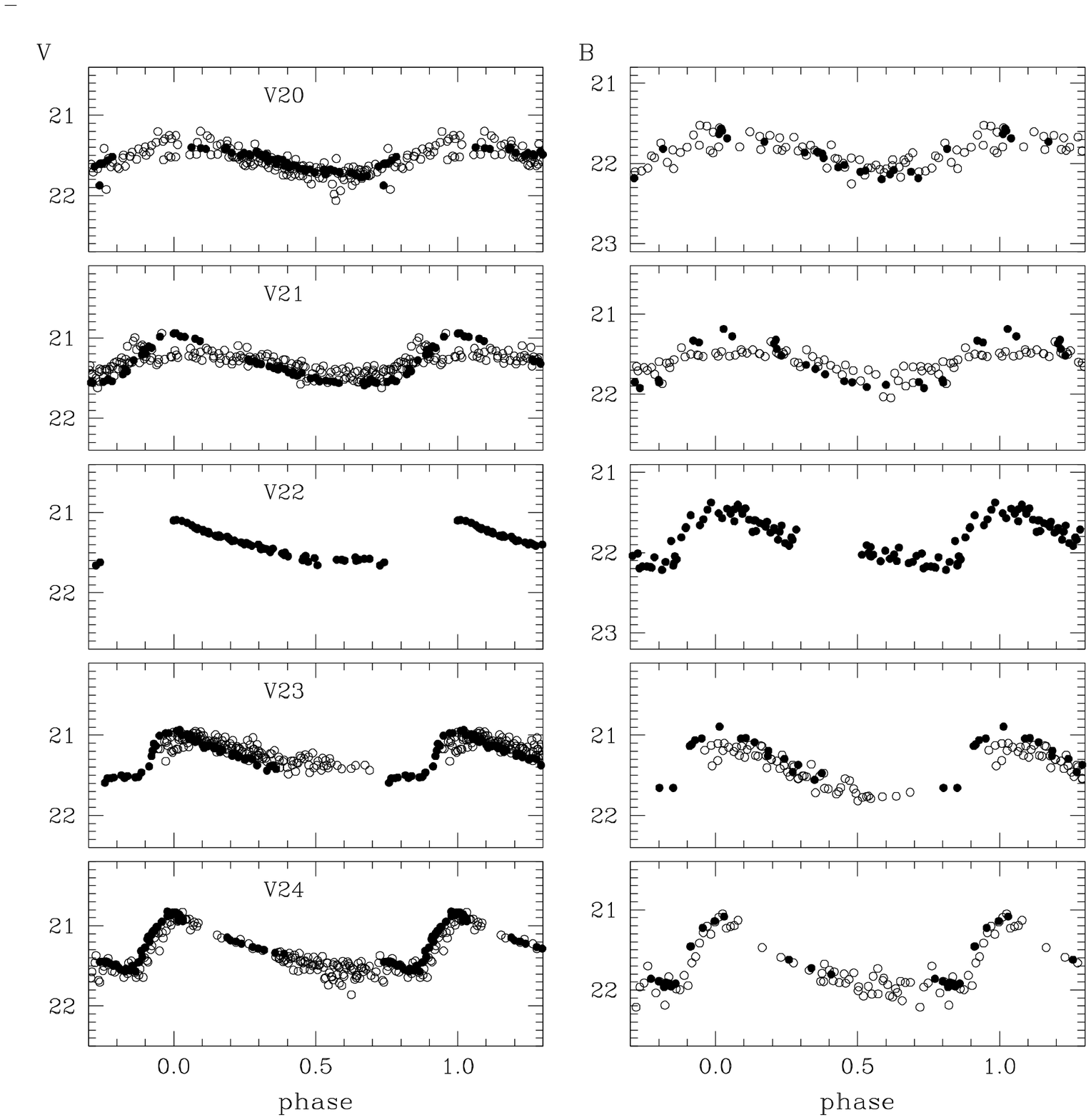}
%\includegraphics[bb=18 144 592 718, width=18cm, draft=true]{difabrizio.figA1_2.jpg}
%%\caption[]{$V$ (left panels), $B$ (right panels) light curves
%%from the combined Magellan plus CTIO datasets.}
\caption{continued}
%\label{f:fig1b}
\end{figure*}

\begin{figure*} 
\figurenum{6}
\includegraphics[width=18cm,bb=40 159 580 703,clip]{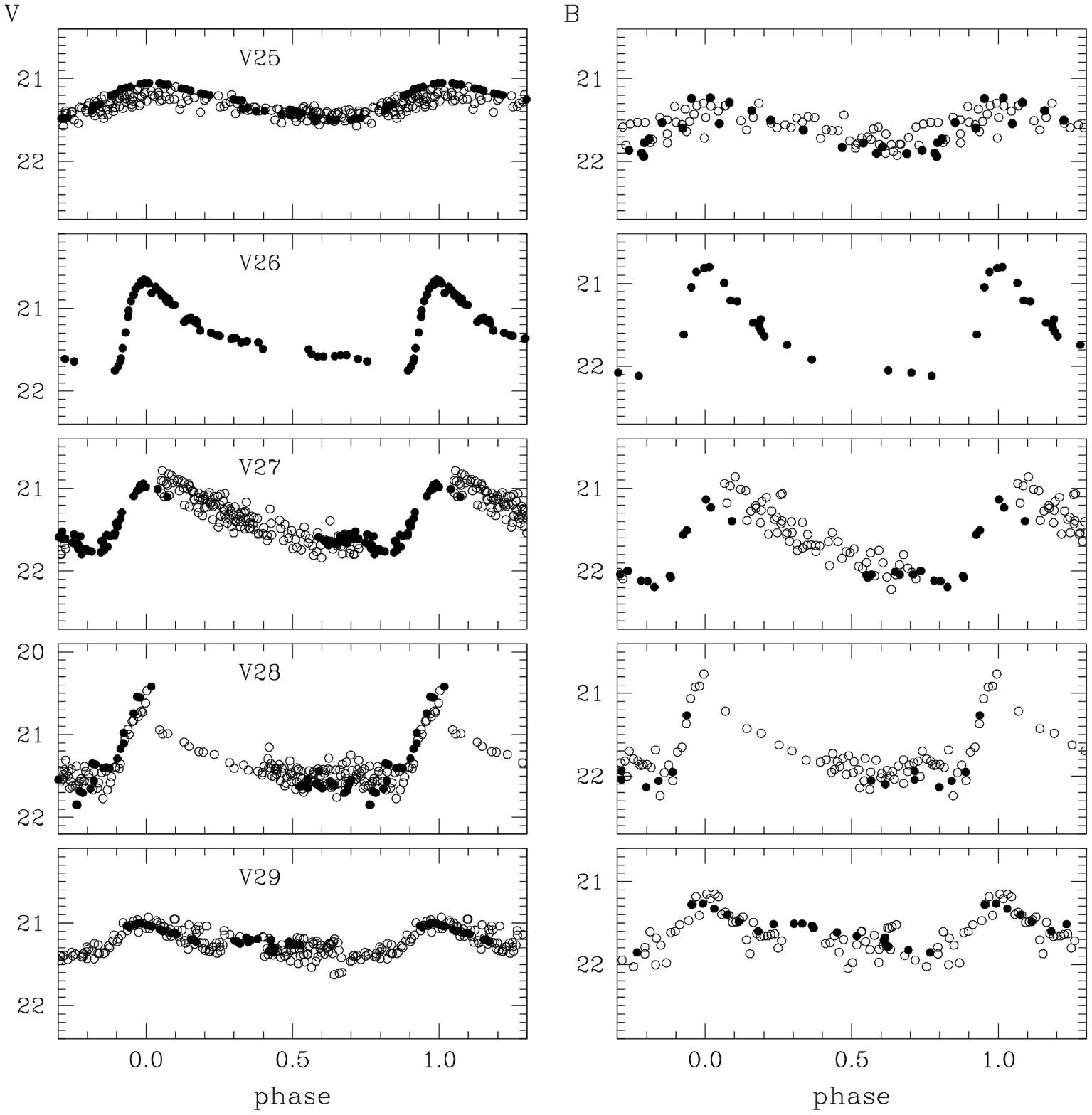}
%\includegraphics[bb=18 144 592 718, width=18cm, draft=true]{difabrizio.figA1_2.jpg}
%%\caption[]{$V$ (left panels), $B$ (right panels) light curves
%%from the combined Magellan plus CTIO datasets.}
\caption{continued}
%\label{f:fig1b}
\end{figure*}

%%\begin{figure}
%%\plotone{c5.ps}
%%\caption{Light curves in differential $V$ flux}
%{f:fig3}
%%\end{figure}

%%\begin{figure}
%%\plotone{c1.ps}
%%\caption{$V$, $B$ light curves.}
%{f:fig3}
%%\end{figure}

%%\begin{figure}
%%\plotone{c2.ps}
%%\caption{$V$, $B$ light curves.}
%{f:fig3}
%%\end{figure}

%%\begin{figure}
%%\plotone{c3.ps}
%%\caption{$V$, $B$ light curves.}
%{f:fig3}
%%\end{figure}

%%\begin{figure}
%%\plotone{c4.ps}
%%\caption{$V$, $B$ light curves.}
%{f:fig3}
%%\end{figure}


\clearpage

\begin{thebibliography}{}

\bibitem[\protect\citeauthoryear{Alard}{1998}]{alard98} Alard, C., \& Lupton, R. H. 1998,  \apj, 503, 325

\bibitem[\protect\citeauthoryear{Alard}{2000}]{alard00} Alard, C. 2000,  A\&AS, 144, 363

\bibitem[\protect\citeauthoryear{Alcock}{2000}]{alcock00} Alcock, C., et al. 2000, \aj, 119, 2194

\bibitem[\protect\citeauthoryear{Ashman}{1994}]{as94} Ashman, K. M., Bird, C. M., \& Zepf, S. E. 1994, \aj, 108, 2348

\bibitem[\protect\citeauthoryear{Barning}{1963}]{barning63} Barning, F. J. M. 1963, Bull. Astron. Inst.
Netherlands, 17, 22

\bibitem[\protect\citeauthoryear{Becker}{2004}]{becker04} Becker, A. C., et al. 2004, \apj, 611, 418

\bibitem[\protect\citeauthoryear{Bellazzini}{2003}]{bellazzini03} Bellazzini, M., Ferraro, F. R., \& 
Ibata, R. 2003, \aj, 125, 188

\bibitem[\protect\citeauthoryear{Bertin}{2002}]{bertin02} Bertin, E., Mellier, Y., Radovich, M., Missonnier, G., 
Didelon, P., \& Morin, B. 2002, in Astronomical Data Analysis Software and Systems XI, eds. D. A. Bohlender,  
D. Durand \& T. H. Handley,  ASP Conf. Ser. 281, 228 

\bibitem[\protect\citeauthoryear{Brown}{1997}]{brown97} Brown, J. A., Wallerstein, G., \& Zucker, D. 1997,
\aj, 114, 180

\bibitem[\protect\citeauthoryear{Buonanno et al.}{1999}]{buonanno99} Buonanno, R., Corsi, C. E., Castellani, M., Marconi, G., Fusi Pecci, F. \& Zinn, R. 1999, \aj, 118, 1671

\bibitem[\protect\citeauthoryear{Buonanno et al.}{1985}]{buonanno85} Buonanno, R., Corsi, C. E., Fusi Pecci, F., Hardy, E., \&  Zinn, R. 1985, \aap, 152, 65

\bibitem[\protect\citeauthoryear{Buonanno et al.}{1998}]{buonanno98} Buonanno, R., Corsi, C. E., Zinn, R., Fusi Pecci, F., Hardy, E., \& Suntzeff, N. B. 1998, \apjl, 501, L33

 \bibitem[\protect\citeauthoryear{Cacciari \& Clementini}{2003}]{cc03}
 Cacciari, C., \& Clementini, G.\ 2003, in Stellar Candles for the
 Extragalactic Distance Scale, ed. D. Alloin and W. Gieren (Berlin:
 Springer, LNP), Lecture Notes in
 Physics, 635, 105

\bibitem[\protect\citeauthoryear{Cacciari et al.}{2005}]{cacciari05} Cacciari, C., Corwin, T. M., \& Carney, B. W. 2005, \aj, 129, 267

%%\bibitem[\protect\citeauthoryear{Cardelli et al.}{1989}]{cardelli89} Cardelli, J. A., Clayton, G. C. \& Mathis, J. S. 1989, \apj, 345, 245

\bibitem[\protect\citeauthoryear{Catelan}{1992}]{catelan92} Catelan, M. 1992, \aap, 261, 443

\bibitem[\protect\citeauthoryear{Catelan}{1993}]{catelan93} Catelan, M. 1993, \aaps, 98, 547

\bibitem[\protect\citeauthoryear{Catelan}{2004a}]{catelan04a} Catelan, M. 2004a, \apj, 600, 409

\bibitem[\protect\citeauthoryear{Catelan}{2004b}]{catelan04b} Catelan, M. 2004b, 
in Variable Stars in the Local Group, ed. D. W. Kurtz \& K. R. Pollard, ASP ASP Conf. Ser., 113

\bibitem[\protect\citeauthoryear{Catelan}{2005}]{catelan05} Catelan, M. 2005,
in Resolved Stellar Populations, ed. D. Valls-Gabaud \& M. Ch\'avez, in press (astro-ph/0507464) 

\bibitem[\protect\citeauthoryear{Catelan}{2007}]{catelan07} Catelan, M. 2007, 
in Globular Clusters~-- Guides to Galaxies, ed.  D.Geisler \& T. Richtler, in press

\bibitem[Clement \& Rowe(2000)Clement \& Rowe]{cr00}Clement, C. M., \& Rowe, J. 2000, \aj, 120, 2579

\bibitem[\protect\citeauthoryear{Clement et al.}{2001}]{clement01}Clement, C. M., et al. 2001, \aj, 122, 2587

\bibitem[\protect\citeauthoryear{Clementini et al.}{2000}]{clementini00}Clementini, G., 
%Di Tomaso, S., Di Fabrizio, L., Bragaglia, A., Merighi, R.,
%Tosi, M., Carretta, E., Gratton, R.G., Ivans, I.I., Kinard, A., Marconi, M., Smith, H.A.,  Wilheim, R., Woodruff, T. 
%\& Sneden, C. 
et al. 2000, \aj, 120, 2054 

\bibitem[\protect\citeauthoryear{Clementini et al.}{2003}]{clementini03}
 Clementini, G., Gratton, R., Bragaglia, A., Carretta, E., Di Fabrizio, L., \& Maio, M.\ 2003, \aj,
 125,1309

\bibitem[\protect\citeauthoryear{Clementini et al.}{2006}]{clementini06}
 Clementini, G., et al. 2006, Mem.S.A.It., 77, 249
%\bibitem[\protect\citeauthoryear{Clementini et al.}{2005}]{clementini05} 
%Clementini, G., Ripepi, V., Bragaglia, A., Martinez Fiorenzano, A., Held, E.V., 
%Gratton, R.G. 2005, \mnras, submitted, (astro-ph/0506206, C05)

\bibitem[\protect\citeauthoryear{Demers et al.}{1990}]{demers90} Demers, S., Kunkel, W. E.,
\& Grodin, L. 1990, \pasp, 102, 632

\bibitem[\protect\citeauthoryear{Di Fabrizio et al.}{1999}]{df99} Di Fabrizio, L.,
1999, {\emph {Laurea Thesis}}, Universit\`a degli Studi di Bologna

\bibitem[\protect\citeauthoryear{Eggen et al.}{1962}]{els} Eggen, O. J., Lynden-Bell, D., 
\& Sandage, A. R. 1962, \apj, 136, 748

\bibitem[\protect\citeauthoryear{Gratton et al.}{2004}]{gratton04} Gratton, R. G., 
Bragaglia, A., Clementini, G., Carretta, E., Di Fabrizio, L., Maio, M., \&
Taribello, E. 2004, \aap, 421, 937

\bibitem[\protect\citeauthoryear{Greco et al.}{2007}]{gre07} Greco, C., et al. 2007, 
in Resolved Stellar Populations, eds. D. Valls-Gabaud \& M. Chavez, 
in press (astro-ph/0507244)

\bibitem[\protect\citeauthoryear{Hardy}{2002}]{h02} Hardy, E. 2002, in IAU Symp. 207, 
Extragalactic Star Clusters, ed. D. Geisler, E. Grebel, \& D. Minniti (San
Francisco:ASP), 62

\bibitem[\protect\citeauthoryear{Harris}{1996}]{harris96} Harris, W. E. 1996, \aj, 112, 1487

\bibitem[\protect\citeauthoryear{Hodge}{1961}]{h61}Hodge, P. W. 1961, \aj, 66, 83

\bibitem[\protect\citeauthoryear{Hodge}{1965}]{h65}Hodge, P. W. 1965, \apj, 141, 308

\bibitem[\protect\citeauthoryear{Hodge}{1969}]{h69}Hodge P. W. 1969, \aj, 720, 249

\bibitem[\protect\citeauthoryear{jurcsik}{1995}]{ju95} Jurcsik, J. 1995, AcA, 45, 653

\bibitem[\protect\citeauthoryear{Jurcsik \& kovacs}{1996}]{jk96} Jurcsik, J., \& Kov\'acs, G. 1996, \aap, 312, 111 

\bibitem[\protect\citeauthoryear{Kaluzny et al.}{1995}]{jkea95} Kaluzny, J., Krzeminski, W., \& Mazur, B. 1995, \aj, 110, 2206

%%\bibitem[\protect\citeauthoryear{kovacs \& Jurcsik}{1997}]{kj97} Kov\'acs, G., \& Jurcsik, J. 1997,
%%A\&A, 322, 218

\bibitem[\protect\citeauthoryear{kovacs \& Kanbur}{1998}]{kk98} Kov\'acs, G., \& Kanbur, S.M., 1998, \mnras, 295, 834 

%%\bibitem[\protect\citeauthoryear{kovacs \& Walker}{2001}]{kw01} Kov\'acs, G., \& Walker, A.R, 2001, \aap, 371, 579 

\bibitem[\protect\citeauthoryear{Landolt}{1992}]{landolt92} Landolt, A. U. 1992, \aj, 104, 340

\bibitem[\protect\citeauthoryear{Lomb}{1976}]{lomb76}Lomb, N. R. 1976, \apss, 39, 447

\bibitem[\protect\citeauthoryear{Mackey \& Gilmore}{2003b}]{mg03b}Mackey, A. D., \& Gilmore, G. F. 2003b, \mnras, 340,
175 

\bibitem[\protect\citeauthoryear{Mackey \& Gilmore}{2003a}]{mg03a}Mackey, A. D., \& Gilmore, G. F. 2003a, \mnras, 345,
747 

\bibitem[\protect\citeauthoryear{Miceli et al.}{2007}]{miceli07}Miceli, A., Rest, A., Stubbs, C.W., Hawley, S.L.,
Cook, K.H., Magnier, E.A., Krisciunas, K., Bowell, E., \& Koehn, B. 2007, arXiv:0706.1583

\bibitem[\protect\citeauthoryear{Oosterhoff}{1939}]{oosterhoff39} Oosterhoff, P. Th. 1939, 
Observatory, 62, 104

\bibitem[\protect\citeauthoryear{Phillips}{1995}]{ph95} Phillips, A. C. \& Davis, L. E. 1995, 
in  Astronomical Data Analysis Software and Systems IV, eds. R. A. Shaw, H. E. Payne \& J. J. E. Hayes, 
ASP Conf. Ser., 77, 297 

\bibitem[\protect\citeauthoryear{Poretti}{2001}]{po01} Poretti, E. 2001, \aap, 371, 986

\bibitem[\protect\citeauthoryear{Poretti et al.}{2006}]{po06} Poretti, E. et al. 2006, Mem.S.A.It., 77, 219

\bibitem[\protect\citeauthoryear{Poretti et al.}{2007}]{po07} Poretti, E., Dell'Arciprete, L., Greco, C., Clementini, G.,
Held, E.V., Pasinetti, L.E., Gullieuszik, M., Maio, M., \& Rizzi, L. 2007, Communication in Asteroseismology, 
in press

\bibitem[Pritzl et al. (2002)Pritzl]{pr02}Pritzl, B. J., Smith, H. A., Catelan, M., \& Sweigart, A. V.  2002, 
\aj, 124, 949; erratum: 2003, \aj, 125, 2752

%%%\bibitem[\protect\citeauthoryear{Pritzl et al.}{2003}]{pritzl03} Pritzl, B. J., Smith, H. A., 
%%%%Stetson, P. B., Catelan, M., Sweigart, A. V., Layden, A. C., \& Rich, R. M. 2003, \apj, 596, L47

\bibitem[2003]{pr03}Pritzl, B. J., Smith, H. A., Stetson, P. B., Catelan, M., Sweigart, A. V., Layden, 
A. C. \& Rich, R. M. 2003, \aj, 126, 1381 

\bibitem[\protect\citeauthoryear{Rest et al.}{2005}]{rest05} Rest, A., 
%Stubbs, C., Becker, A.C., Miknaitis, G.A., Miceli, A., Covarrubias, R., Hawley,S.L., Smith, R.C.,
%Suntzeff, N.B., Olsen, K., Prieto, J.L., Hiriart, R., Welch, D.L., Cook, K.H., Nikolaev, S., Huber, M., Challis, P., 
%Keller, S.C. \& Schmidt, B.P. 
et al. 2005, \apj, 634, 1103

\bibitem[\protect\citeauthoryear{Sandage}{1990}]{sa90} Sandage, A. 1990, JRASC, 84, 70

\bibitem[\protect\citeauthoryear{Sandage}{1993}]{sa93} Sandage, A. 1993, \aj, 106, 687

\bibitem[\protect\citeauthoryear{Sandage}{1996}]{sa96} Sandage, A. 2006, \aj, 131, 1750 

\bibitem[\protect\citeauthoryear{Saviane et al.}{2000}]{saviane00} Saviane, I., Held, E. V., \& Bertelli, G. 2000,
\aap, 355, 56

\bibitem[\protect\citeauthoryear{Scargle}{1982}]{scargle82} Scargle, J. D. 1982, \apj, 263, 835

\bibitem[\protect\citeauthoryear{Schechter}{1993}]{sc93} Schechter, P. L., Mateo, M., \& Saha, A., 1993, \pasp, 105, 1342

\bibitem[\protect\citeauthoryear{Searle \& Zinn}{1978}]{sz} Searle, L., \& Zinn, R. 1978, \apj, 225, 357

\bibitem[Soszy\'{n}ski et al.(2003)]{isea03} Soszy\'{n}ski, I., et al. 2003, AcA, 53, 93


\bibitem[\protect\citeauthoryear{Stetson  et al.}{1994}]{st94} Stetson, P. B. 1994, \pasp, 106, 250

\bibitem[\protect\citeauthoryear{Stetson et al.}{1996}]{st96} Stetson, P. B. 1996, \emph{User's Manual for DAOPHOT II}

\bibitem[\protect\citeauthoryear{Strader et al.}{2003}]{strader03} Strader, J., Brodie, J. P., Forbes, D. A., 
Beasley, M. A., \& Huchra, J. P. 2003, \aj, 125, 1291

\bibitem[\protect\citeauthoryear{van den Bergh}{1993}]{vdb93} van den Bergh, S. 1993, \mnras, 262, 588

\bibitem[\protect\citeauthoryear{Yoon \& Lee}{2002}]{yl02} Yoon, S-J, Lee, Y-W, 2002, Science, 297, 578

\bibitem[\protect\citeauthoryear{Vanicek}{1971}]{pv71}Vanicek, P. 1971, \apss, 12, 10

\bibitem[\protect\citeauthoryear{Zinn \& West}{1984}]{zw84} Zinn, R., \& West, M. J. 1984,
\apjs, 55, 45

%\bibitem[Baade \& Hubble(1939)Baade \& Hubble]{bh39}Baade, W. \& Hubble, E. 1939, \pasp, 51, 40

%\bibitem[Bersier \& Wood(2002)Bersier \& Wood]{bw02}Bersier, D. \& Wood, P. R. 2002, \aj, 123, 840
%\bibitem[Buonanno et al. (1985)Buonanno, Corsi, \& Fusi Pecci]{buonanno85} Buonanno, R., Corsi, C. E.,Fusi Pecci, F. \& Hardy, E. 1985,  A\&A, 152, 65
%\bibitem[Buonanno et al. (1998)Buonanno]{buonanno98}Buonanno, R., Corsi, C. E., Zinn, R., Fusi Pecci, F., Hardy, E. \& Suntzeff, N. B. 1998, ApJ, 501, L33
%%\bibitem[Burnstein \& Heiles (1982)]{bh82}Burnstein, D. \& Heiles, C. 1982, AJ, 87, 1165
%\bibitem[Castellani et al.(2003)Castellani, Caputo, \& Castellani]{castellani03}Castellani, M., Caputo, F., \& Castellani, V. 2003,  A\&A, 410, 871
%\bibitem[Catelan(2004)]{c04}
%  Catelan, M. 2004, in Variable Stars in the Local Group, ASP Conf. Ser.,
%  310, ed. D. W. Kurtz \& K. R. Pollard (San Francisco: ASP), 113
%\bibitem[Clement \& Rowe(2000)Clement \& Rowe]{cr00}Clement, C. M. \& Rowe, J. 2000, AJ, 120, 2579
%\bibitem[Clementini et al.(2004)Clementini]{clementini04}
% Clementini, G. et al. 2004, in Variable Stars in the Local Group, ASP Conf. Ser.,
%  310, ed. D. W. Kurtz \& K. R. Pollard (San Francisco: ASP), 60
%\bibitem[Gratton et al.(2003)Gratton]{g03}Gratton, R.G., Bragaglia, A., Carretta, E., Clementini, G., Desidera, S., Grundahl, F., Lucatello, S. 2003, A\&A , 408, 529
%\bibitem[Gratton et al.(2004)Gratton]{g04}Gratton, R.G., Bragaglia, A.,  Clementini, G., Carretta, E., Di Fabrizio, L., Maio, M., Taribello, E. 2004, A\&A, 421, 937
%\bibitem[Held et al. (2001)Held]{h01}Held, E. V., Clementini, G., Rizzi, L., Momany, Y., Saviane, I., Di Fabrizio, L. 2001, ApJ, 562, L39
%\bibitem[Pritzl et al. (2002)Pritzl]{pr02}Pritzl, B. J., Armandroff, T. E., Jacoby, G. H. \& da Costa, G. S. 2002, AJ, 124, 949
%\bibitem[Sandage (1993)Sandage]{sandage03}Sandage, A. 1993, AJ, 106, 687
%\bibitem[Saviane et al. (2000)Saviane, Held, \& Bertelli]{saviane00} Saviane, I., Held, E.V., \& Bertelli, G. 2000,  A\&AS, 355, 56
%\bibitem[Soszynski et al. (2003)Soszynski]{so03}Soszynski, I., Udalski, A., Szymanski, M., Kubiak, M., Pietrzynski, G., Wozniak, P., Zebrun, K., Szewczyk, 
%O., Wyrzykowski, L. et al. 2003, , A\&A, 53, 93
%\bibitem[Stetson et al. (1998)Stetson, Hesser, \& Smecker-Hane]{stetson98} Stetson, P. B., Hesser, J., \& Smecker-Hane, T. 1998, PASP50, 110, 533
\end{thebibliography}
\end{document}